\documentclass[%
 twocolumn, 
 superscriptaddress,
 amsmath,amssymb,
 aps, physrev,
]{revtex4-2}
\usepackage{graphicx}% Include figure files
\usepackage{dcolumn}% Align table columns on decimal point
\usepackage{bm}% bold math
\usepackage{xcolor}
\usepackage{comment}
\begin{document}

%\preprint{APS/123-QED}

%Title of paper
\title{Intrinsic back-switching phenomenon in SOT-MRAM devices}

% repeat the \author .. \affiliation  etc. as needed
% \email, \thanks, \homepage, \altaffiliation all apply to the current
% author. Explanatory text should go in the []'s, actual e-mail
% address or url should go in the {}'s for \email and \homepage.
% Please use the appropriate macro foreach each type of information

% \affiliation command applies to all authors since the last
% \affiliation command. The \affiliation command should follow the
% other information
% \affiliation can be followed by \email, \homepage, \thanks as well.
\author{Kuldeep Ray}
%\email[]{Your e-mail address}
%\homepage[]{Your web page}
%\thanks{}
%\altaffiliation{}
\email{Contact author: kuldeepray10@gmail.com}
\affiliation{Univ. Grenoble Alpes, CEA, CNRS, Grenoble-INP, SPINTEC, 38000 Grenoble, France}
\affiliation{Antaios, 38240 Meylan, France}

\author{Jérémie Vigier}
\affiliation{Univ. Grenoble Alpes, CEA, CNRS, Grenoble-INP, SPINTEC, 38000 Grenoble, France}
\affiliation{Antaios, 38240 Meylan, France}

\author{Perrine Usé}
\affiliation{Antaios, 38240 Meylan, France}

\author{Sylvain Martin}
\affiliation{Antaios, 38240 Meylan, France}

\author{Nicolas Lefoulon}
\affiliation{Antaios, 38240 Meylan, France}

\author{Chloé Bouard}
\affiliation{Antaios, 38240 Meylan, France}

\author{Marc Drouard}
\affiliation{Antaios, 38240 Meylan, France}

\author{Gilles Gaudin}
\email{Contact author: gilles.gaudin@cea.fr}
\affiliation{Univ. Grenoble Alpes, CEA, CNRS, Grenoble-INP, SPINTEC, 38000 Grenoble, France}

%Collaboration name if desired (requires use of superscriptaddress
%option in \documentclass). \noaffiliation is required (may also be
%used with the \author command).
%\collaboration can be followed by \email, \homepage, \thanks as well.
%\collaboration{}
%\noaffiliation

\date{\today}

\begin{abstract}
The writing process of SOT-MRAMs is considered deterministic when additional symmetry-breaking factors, such as the application of an external magnetic field aligned with the current, are present. Notably, the write probability exhibits a unique behavior as a function of the current: it drops to zero at high currents or even oscillates with the current. This phenomenon is attributed to back-switching, an intrinsic effect of magnetization reversal driven by spin-orbit torques. A systematic investigation of this back-switching phenomenon is conducted on sub-100~nm CoFeB magnetic pillars positioned at the center of $\beta$-W Hall crosses. Using a statistical approach, the study examines the impact of various parameters, including the amplitude of current pulses and the application of magnetic fields in different directions. The findings reveal that the back-switching phenomenon is not statistically random. Macrospin simulations, employing realistic magnetic parameter values, accurately replicate the experimental observations and provide insights into the underlying mechanisms of back-switching. These simulations also explore strategies to mitigate the phenomenon, such as optimizing the shape of the writing pulses. Applying this approach to complete SOT-MRAM single cells achieves a write error rate below $2\times10^{{-6}}$, demonstrating the effectiveness of this strategy in expanding the operational current range for write operations in SOT-MRAMs.
\end{abstract}

% insert suggested keywords - APS authors don't need to do this
%\keywords{}

%\maketitle must follow title, authors, abstract, and keywords
\maketitle

% body of paper here - Use proper section commands
% References should be done using the \cite, \ref, and \label commands
\section{Introduction}

Magnetic random-access memories (MRAMs) are considered one of the most promising replacements for volatile CMOS memory technologies across all levels of the memory hierarchy. Among the latest generations of MRAMs, spin-orbit torque MRAMs (SOT-MRAMs) are particularly attractive for replacing memories closest to the processing units, as they combine high endurance with sub-nanosecond switching times \cite{prenat_ultra-fast_2016, shao_roadmap_2021}. Nevertheless, while spin transfer torque MRAMs (STT-MRAMs) are commercially available \cite{noauthor_spin-transfer_nodate}, SOT-MRAMs must overcome several challenges before they can be adopted by the industry \cite{shao_roadmap_2021}. One of these major challenges is to achieve deterministic switching, meaning reliably controlling the result of the write operation, over a wide current range. The key metric for deterministic switching is the write error rate, which measures the probability of getting the expected final state after a write operation.

In SOT-MRAM technology with perpendicular magnetic anisotropy, which is the most promising solution in terms of scaling, density and data retention, magnetization switching by means of spin-orbit current alone is not deterministic. Indeed, by symmetry, the probabilities of obtaining an "up" or "down" magnetization are equal. An additional factor to break this symmetry is required. In early implementations, this breaking of symmetry was induced by the application of an in-plane magnetic field, collinear with the current but external to the system \cite{miron_perpendicular_2011, cubukcu_spin-orbit_2014, liu_spin-torque_2012}. As this solution is not suitable for embedded applications, a number of propositions have been published in recent years, aimed at developing deterministic and integrated "field-free" SOT-MRAM cells, based on tilted anisotropy \cite{you_switching_2015}, combination of SOT and STT effects \cite{cai_sub-ns_2021}, exchange bias \cite{van_den_brink_field-free_2016}, embedded magnet in the material stack \cite{garello_manufacturable_2019} and unconventional torques from new materials \cite{kao_deterministic_2022} (for a more complete list, please consult review articles such as \cite{krizakova_spin-orbit_2022}).

\begin{figure}
\includegraphics[width=\linewidth]{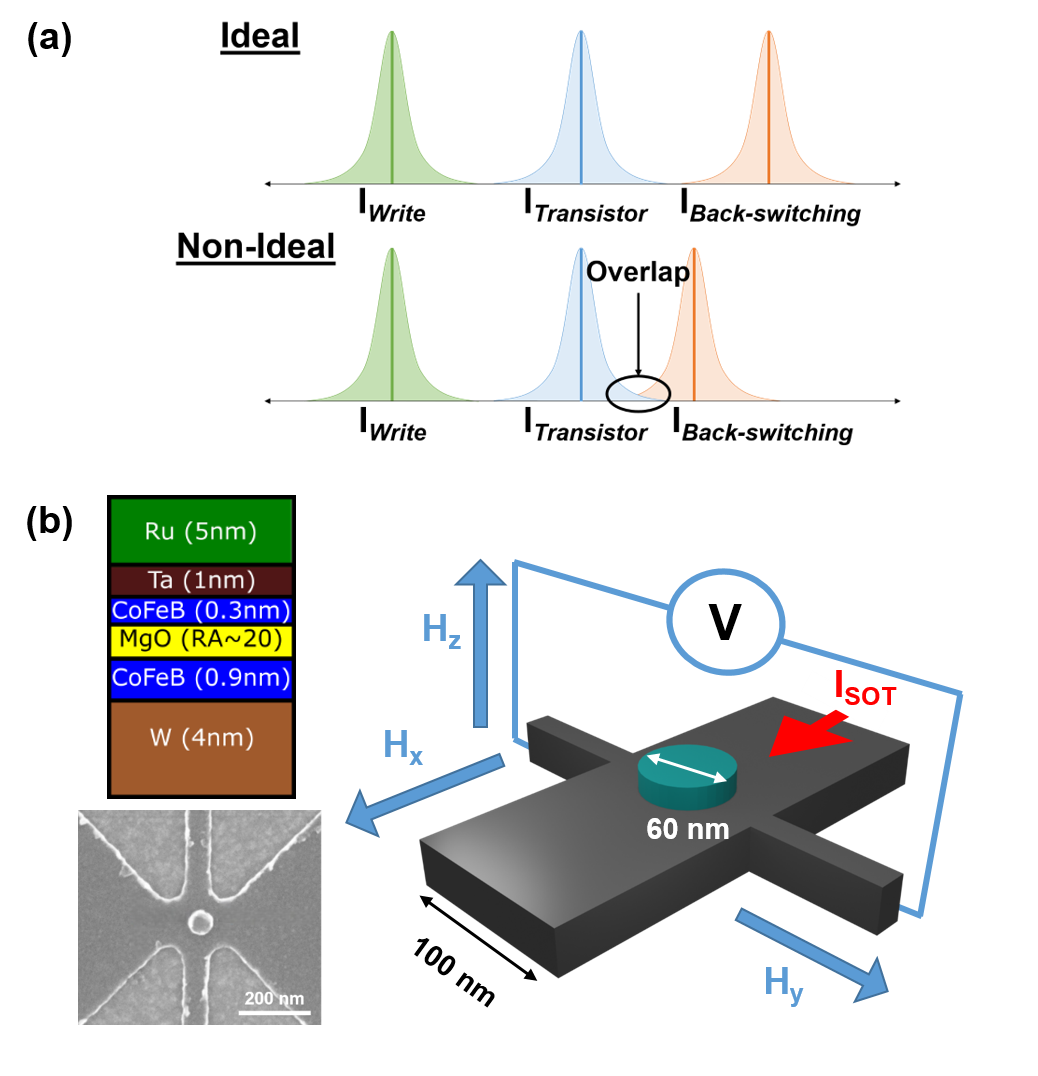}
\caption{\label{fig:device_dimension}(a) To design a memory array, the statistical variation in device properties must be taken into account. To ensure reliable writing, the distribution of the current supplied by the write transistor (I$_{\text{Transistor}}$) should be higher than the distribution of SOT-MRAM write current (I$_{\text{Write}}$) and should not overlap with the distribution of the current inducing a back-switching phenomenon (I$_{\text{Back-switching}}$). Ideally, I$_{\text{Transistor}}$ should be in the middle of I$_{\text{Write}}$ and I$_{\text{Back-switching}}$. (b) A schematic representation of the studied magnetic stack and the device used for electrical measurements is shown: a magnetic pillar patterned onto a $\beta$-tungsten Hall cross (see the representative SEM image). The applied writing current I$_{\text{SOT}}$ is injected through the longitudinal arms of the Hall cross, while the anomalous Hall voltage is measured across the transverse arms. H$_{\text{X}}$, H$_{\text{Y}}$, and H$_{\text{Z}}$ denote the applied magnetic field components along the respective axes.}
\end{figure}

However, the presence of symmetry breaking does not guarantee a perfectly deterministic reversal of magnetization, and consequently a deterministic writing. It has been observed that increasing the write current beyond the threshold for reversal leads, as expected, to an increase in the probability of magnetization switching. However, if the current is increased further, the probability then decreases and may even oscillate as a function of the applied current \cite{lee_oscillatory_2018, yoon_anomalous_2017, decker_time_2017}. Similar back-switching (BSW) phenomenon has previously been observed and studied in STT-MRAMs \cite{min_back-hopping_2009}. One origin of this phenomenon has been attributed to the instability of the reference layer magnetization in magnetic tunnel junctions (MTJ), an instability that can even lead to the reversal of this magnetization \cite{kim_experimental_2016, devolder_back_2020}. In SOT-MRAMs, magnetization is switched by means of a current injected in the plane of the layers \cite{miron_perpendicular_2011, liu_spin-torque_2012}, without the need for an MTJ-type structure to obtain a spin-polarized current. Observation of the back-switching phenomenon in simple structures, such as non-magnetic/ferromagnetic/insulator trilayer stacks, demonstrates that in SOT-MRAM this BSW is intrinsic to the magnetization dynamics induced by spin-orbit torques and not related to the dynamics of the reference layer. This BSW phenomenon is a challenge for the efficient design of memories based on SOT-MRAMs (Fig.~\ref{fig:device_dimension}~(a)). Its study is essential to our understanding of the magnetization dynamics induced by spin-orbit torques, and for finding effective means of mitigating BSW in SOT-MRAMs.

Despite its importance, relatively few studies have been devoted to it. Most are numerical studies based on macrospin simulations. The existence of this BSW phenomenon was reported as early as the first macrospin SOT switching simulations, whether with or without the Field-Like (FL) term \cite{lee_threshold_2013, park_macrospin_2014}. The importance and the role of the FL term on the BSW phenomenon were then highlighted \cite{legrand_coherent_2015, taniguchi_theoretical_2019, zhu_threshold_2020}. The existence of this BSW phenomenon has been explained by energy arguments: at the end of the write current pulse, if the energy of the system is higher than the saddle point energy, the magnetization will relax in either a first potential well corresponding to a first direction of magnetization or in the other corresponding to the opposite state of magnetization \cite{park_macrospin_2014, taniguchi_theoretical_2019}. This explains the high sensitivity of the phenomenon to damping, as reported in particular by \cite{lee_threshold_2013, park_macrospin_2014, taniguchi_reduction_2020}. Experimentally, this phenomenon has been reported in some publications - for example, in Pt/Co/AlO$_{\text{X}}$ layers, where the magnetization is reported to switch back to its original orientation above a certain threshold of applied planar magnetic field \cite{decker_time_2017}; or in Ta/CoFeB/MgO-based magnetic tunnel junctions, where a decrease in the probability of the magnetization switching is observed when the write current is increased \cite{krizakova_tailoring_2022}. Only two publications from the same group report a more complete study of this phenomenon, combining experiments and simulations \cite{yoon_anomalous_2017, lee_oscillatory_2018}. Both studies were carried out on micron-sized devices patterned from a Ta/CoFeB/MgO based stack. Time-resolved MOKE measurements show a decrease in the probability of switching as the writing current increases. Using analytical modeling and micromagnetic simulations, these measurements were explained by a BSW phenomenon due to the reflection of the magnetic domain walls (DW) on the edges of the device. This reflection can only occur for a negative sign of the ratio of the FL and Damping-Like (DL) torques \cite{yoon_anomalous_2017}. In a subsequent study, based on electrical measurements on Hall crosses supporting a magnetic pillar, the authors observed an oscillatory and deterministic behavior of the magnetization switching while increasing the amplitude of nanosecond write pulses \cite{lee_oscillatory_2018}. 

In this work, we present a systematic study of the back-switching phenomenon using a statistical approach. The measurements were carried out on sub-100nm CoFeB magnetic pillars at the center of $\beta$-W Hall crosses. The error rate of the write operation was measured as a function of various parameters such as the amplitude of the write current pulse and an external magnetic field applied along different directions. We show that macrospin simulations, using realistic values for the different magnetic parameters, reproduce the experimental observations well. These macrospin simulations open the door to the development of compact models that take this BSW phenomenon into account. These numerical simulations give us an insight into the mechanism behind the BSW phenomenon. They also allow us to study ways of reducing this phenomenon, such as using stronger damping, tuning the FL/DL ratio or, following previous work from \cite{zhu_threshold_2020} controlling the shape of the writing pulses. We test the latter solution and show experimentally that this strategy does indeed mitigate back-switching and increases the current range for write operations in SOT-MRAMs.

\section{Methodology}

The measurements were carried out on a stack constituting the typical storage layer of a SOT-MRAM: $\beta$-W(4)/CoFeB(0.9)/MgO(RA$\sim$20~$\Omega\mu$m$^2$)/CoFeB(0.3)/ Ta(1)/Ru(5) where the thicknesses are expressed in nanometers. The top 0.3~nm CoFeB layer is non-magnetic and is used to provide appropriate environment to the MgO layer during annealing. The layers were sputter deposited and annealed at 350\textdegree~C for 30~minutes unless otherwise specified in the text. This stack was then patterned as a CoFeB/MgO/CoFeB/Ta/Ru pillar at the center of a $\beta$-W Hall cross by successive steps of electron lithography and ion beam etching. The studies reported here, except for certain cases specified in the text, were carried out on pillars with a real diameter of around 60~nm sitting on a 100~nm wide track. Fig.~\ref{fig:device_dimension}~(b) presents the magnetic stack, a representative SEM image of a studied device, and a schematic of the measurement setup.

\begin{figure}
\includegraphics[width=\linewidth]{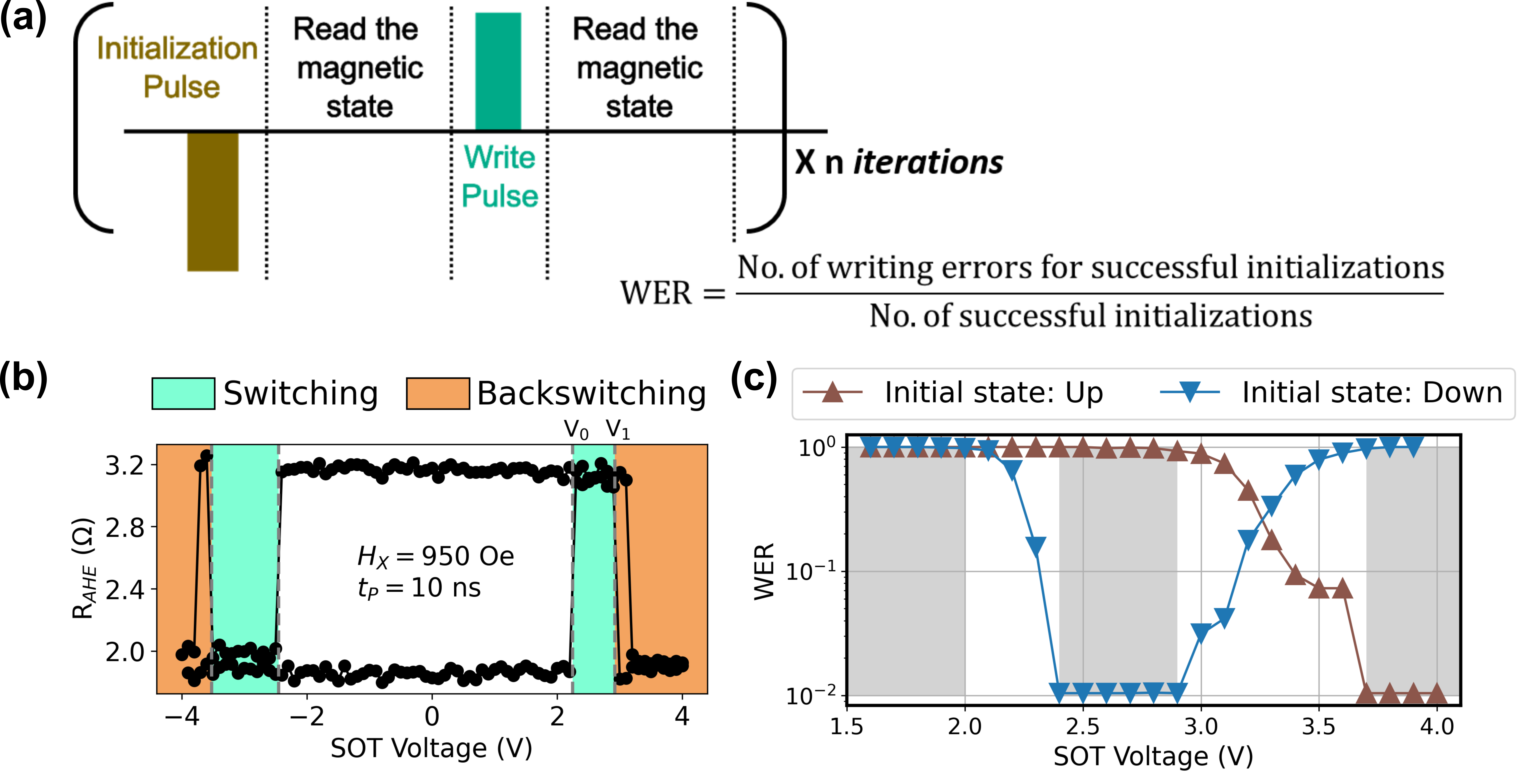}
\caption{\label{fig:wer_scheme}(a) WER measurement scheme and definition. (b)~Typical hysteresis curve for current-induced magnetization reversal. The current pulse is 10~ns long, with a rise/fall time of 2~ns, and a permanent field (H$_{\text{X}}$) of 950~Oe is applied. The areas colored green (respectively red) define the current ranges for which switching (respectively back-switching) is observed. V$_0$ and V$_1$ are the positive voltages that define the boundaries of these areas (see text). (c)~Corresponding WER measurement curve where larger voltages were applied. A deterministic back-switching is observed while increasing the applied voltage.}
\end{figure}

The study is based on statistical measurements of Write Error Rate (WER), which is a crucial metric for qualifying memory devices. Typically, while this WER is very low in SRAM memories, it is much higher in MRAMs. However, to operate as a working memory, the WER of a single bit should typically be less than 10$^{{-9}}$ if the chip incorporates error correction code (ECC), or less than 10$^{{-18}}$ without this error correction \cite{khvalkovskiy_basic_2013}. Fig.~\ref{fig:wer_scheme}~(a) shows the WER measurement scheme that has been used and its definition: the memory point is initialized to a defined state and this state is read to verify that initialization was successful. A pulse to write the opposite state, of variable amplitude, is then applied and the state of the memory point is read again. WER is defined as the ratio of the number of write errors after a successful initialization to the number of successful initializations. Note that a WER of 0.5 corresponds to an equal probability of obtaining the two states of opposite magnetization.

\section{Experimental Results}

Magnetic characterization of the samples confirms perpendicular anisotropy, as evidenced by square hysteresis loops, shown in \cite{supplementary}. The coercive field is approximately 750~Oe, while the anisotropy field~(H$_{\text{k}}$) and the thermal stability factor~($\Delta$), extracted using the switching field distribution method \cite{tillie_data_2016, feng_sweep-rate-dependent_2004}, are estimated to be around H$_{\text{k}}$=2600~Oe and $\Delta$=45, respectively \cite{supplementary}. A typical hysteresis curve for current induced magnetization reversal is shown in Fig.~\ref{fig:wer_scheme}~(b). The areas colored in green represent known areas of magnetization reversal into the opposite state. However, as the amplitude of the applied voltage increases, a new behavior appears, represented by the areas colored in red. The final state of magnetization is no longer fixed, opposite to the initial state, but begins to fluctuate in an apparently random manner between these two states. This fluctuation occurs before eventually switching back to the initial state, as observed for positive voltages. This behavior is not observed for negative voltages, likely due to a slight asymmetry in the experimental setup, as discussed below.

The corresponding WER measurement is shown in Fig.~\ref{fig:wer_scheme}~(c) for positive voltages. This curve was obtained with 100 write/read cycles, which explains the saturation at WER$=10^{{-2}}$, representing in this case an absence of error. Starting from a down state and applying a positive voltage, the WER of the up state is measured. We first observe a zone without switching when the voltage is below a first threshold V$_0$. Then when the voltage is increased above V$_0$, we observe that, after a transition zone, the magnetization switches to the up state without error, corresponding to the green zone in Fig.~\ref{fig:wer_scheme}~(b). When the voltage increases further, there is another threshold V$_1$ at which the error rate starts increasing and approaches the maximum value 1, which corresponds to the magnetization switching back to its initial orientation, in this case down, without any error. This behavior is repeated when the magnetization is initialized in the opposite up state and by measuring the WER of the down state. By applying a positive voltage below V$_1$, the up state remains stable while above V$_1$, the magnetization switches to the down state. This shows that the BSW is not random but, on the contrary, deterministic: after being reversed to the orientation opposing the initial state, the magnetization can be switched back to its initial orientation in a deterministic way, as shown by the low error rate obtained (cf Fig.~\ref{fig:wer_scheme}~(c)). This behavior is similar to that reported by Lee \textit{et al.} \cite{lee_oscillatory_2018}.

\begin{figure*}
\includegraphics[width=0.83\linewidth]{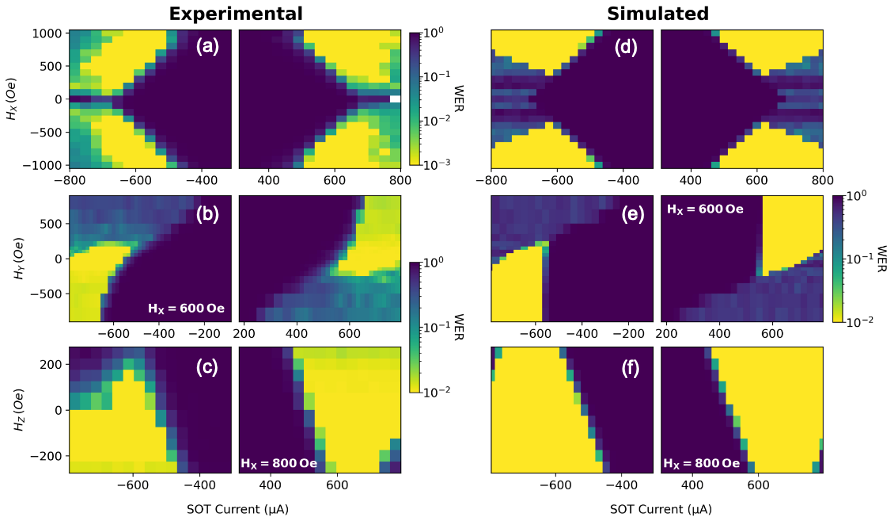}
\caption{\label{fig:wer_colormaps}Experimental and simulated WER color maps as a function of applied SOT current using 5 ns-long pulses with 70 ps rise/fall times. The experimental results are shown in (a) for a variable H$_{\text{X}}$ field, (b) for a variable H$_{\text{Y}}$ field in the presence of a fixed H$_{\text{X}}$ field of 600~Oe, and (c) for a variable H$_{\text{Z}}$ field and a fixed H$_{\text{X}}$ field of 800~Oe. Simulation results are shown opposite for the same applied fields (d), (e) and (f). These WER maps were obtained using 100 write/read iterations. The blue color, corresponding to a value of 1, indicates an error for each of the procedures, and the yellow color indicates no writing error.}
\end{figure*}

In order to characterize this BSW, we systematically measured the WER as a function of current with a magnetic field applied in each of the three directions in space. Fig.~\ref{fig:wer_colormaps} shows the WER maps obtained for the same device. The first thing to notice is that the different maps are not exactly symmetrical for positive and negative currents. This is due to the presence of a small positive DC bias current, I$_{\text{read}} = 10$~µA injected continuously into the device to measure its AHE resistance. However, the disturbance caused is small compared to the switching current ($\sim500$~µA) and does not prevent analysis. The WER measured as a function of an H$_{\text{X}}$ field for a positive current or a negative current (Fig.~\ref{fig:wer_colormaps}~(a)) shows 3 regions. The blue region, corresponding to a maximum value of the WER = 1, is the sub-critical region for which the applied current is less than the magnetization reversal critical current in the presence of the applied H$_{\text{X}}$ field. The yellow region, corresponding to a minimum value of the WER, is the deterministic magnetization reversal region. Finally, the light green-blue region, corresponding to an intermediate value of the WER, is the region where the magnetization begins to switch back towards its initial state. As the H$_{\text{X}}$ field increases, the energy barrier to be crossed for magnetization switching decreases while the initial amplitude of the DL term increases, leading to a decrease in the critical reversal current. This is observed at the boundary between the blue and yellow regions. In addition, this decrease is almost linear with the increase in the applied magnetic field, as anticipated by macrospin modeling \cite{lee_threshold_2013, taniguchi_critical_2015, torrejon_current-driven_2015}. The behavior of the boundary between the yellow and light green-blue regions is more difficult to determine here. For negative currents, the amplitude of the current for which the BSW appears, the BSW threshold current, decreases with an increase in the applied H$_{\text{X}}$ field. For positive currents, no clear trend is observed. These three regions can be found on all the maps shown in Fig.~\ref{fig:wer_colormaps}. 

Fig.~\ref{fig:wer_colormaps}~(b) shows the WER maps as a function of a variable H$_{\text{Y}}$ field and increasing SOT current for an applied H$_{\text{X}}$ field of 600~Oe. When the applied H$_{\text{Y}}$ field is positive or slightly negative (respectively negative or slightly positive) for a positive (respectively negative) applied current, the reversal of the magnetization is stable and deterministic (yellow region). When this applied field changes direction, the WER increases sharply and no deterministic magnetization switching can be obtained. Moreover, for the deterministic reversal regions, the amplitude of the critical reversal current decreases with that of the applied magnetic field, whereas the opposite behavior is observed for the boundary between the switching and back-switching regions. This behavior is consistent with a negative effective field, H$_{\text{FL}}$, produced by a positive injected current: when the applied H$_{\text{Y}}$ field adds to H$_{\text{FL}}$, the energy barrier separating the two magnetization states decreases, leading to a reduction in the critical reversal current \cite{lee_oscillatory_2018, krizakova_tailoring_2022, fan_asymmetric_2019}. However, as the amplitude of the resulting H$_{\text{Y}}$ field increases, the magnetization approaches an in-plane position at the end of the pulse, resulting in a loss of determinism, as explained in the next section. 

Finally, Fig.~\ref{fig:wer_colormaps}~(c) shows the WER maps as a function of a variable H$_{\text{Z}}$ field and SOT current for an applied H$_{\text{X}}$ field of 800~Oe. These maps are in agreement with the hypothesis of an up (respectively down) magnetization orientation stabilized by a positive (respectively negative) magnetic field, H$_{\text{Z}}$. Thus, for positive (respectively negative) currents, the critical reversal current increases (respectively decreases) when the amplitude of the negative magnetic field increases. More (or less) current is needed to stabilize an up (or down) magnetization in the presence of a negative H$_{\text{Z}}$ field. Similar reasoning applies to positive fields. This hypothesis is consistent with the negative sign of the spin Hall effect in W and the presence of a positive H$_{\text{X}}$ field \cite{hao_giant_2015}. We can also see the presence of a blue zone for high positive fields and large negative currents, a zone that also exists for opposite current and H$_{\text{Z}}$ field. During the current pulse, the FL and DL spin orbit torques maintain the magnetization in a dynamic equilibrium whose direction approaches the equator (m$_{\text{Z}}$=0) as the current amplitude increases (see the section on macrospin simulations below). At the end of the current pulse, only the applied magnetic field, H$_{\text{Z}}$, remains, dictating the final orientation of the magnetization.

\section{Discussion and Simulations}

To go further in the analysis of these results, we compared them with the results obtained from simulations. It is known that for pillar diameters such as those we have measured, the reversal of magnetization by SOT occurs through a two-step nucleation-propagation process \cite{mikuszeit_spin-orbit_2015, baumgartner_spatially_2017}. Micromagnetic simulations are therefore the most accurate way of faithfully reproducing magnetization behavior. However, such simulations to calculate WER curves are time-consuming. Therefore, we aimed to reproduce these measurements using macrospin simulations. In this coherent reversal approximation, the magnetization dynamics can be reproduced by the Landau-Lifshitz-Gilbert equation, to which the two terms describing the FL and DL torques of the SOTs \cite{park_macrospin_2014} are added: 

\begin{eqnarray}
\label{eqn:llg} 
\frac{d\mathbf{m}}{dt} &=& -\gamma \mu_0 \left( \mathbf{m} \times \mathbf{H}_{\text{eff}} \right) + \alpha \left( \mathbf{m} \times \frac{d\mathbf{m}}{dt} \right) \nonumber \\
&&  + \gamma \mu_0 \left( \mathbf{m} \times \mathbf{H}_{\text{FL}} \right) + \gamma \mu_0 \left( \mathbf{m} \times \mathbf{H}_{\text{DL}} \right)
\end{eqnarray}

With the effective damping-like field,   $\mathbf{H_{\text{DL}}}~=~\frac{j_{\text{SOT}} \theta_{\text{SH}} \hbar}{2 \mu_0 e t_m M_S} (\mathbf{m} \times \mathbf{u}_{\text{Y}})$ and the effective field-like term, $\mathbf{H_{\text{FL}}}~=~ \beta\frac{j_{\text{SOT}} \theta_{\text{SH}} \hbar}{2 \mu_0 e t_m M_S} \mathbf{u}_{\text{Y}}$ where $e$ is the elementary charge, $t_m$ is the thickness of the free magnetic layer, in this case the CoFeB layer, $M_S$ is its saturation magnetization, $\theta_{\text{SH}}$ is the spin Hall angle, $\beta$ is the FL to DL ratio and $\mathbf{u}_{\text{Y}}$ is a unit vector along Y. The effective field ($\mathbf{H_{\text{eff}}}$) includes any applied external field as well as the anisotropy field whose maximum amplitude is given by H$_{anis}^{eff}~=~\frac{(2K_u^{eff})}{M_S}~-~\mu_0 M_S$. Perpendicular anisotropy is modeled by first-order uniaxial anisotropy $\varepsilon_{\text{anis}}~=~K_u^{\text{eff}}~\sin^2\theta$. In order to simulate WER curves, a random thermal field $H_{\text{th}}~=~\sqrt{\frac{2 \alpha k_B T}{\gamma \mu_0^2 V M_S \Delta t}}~\zeta_{\text{th}}$
 is added to $\mathbf{H_{\text{eff}}}$. k$_B$ is the Boltzmann constant, T~=~300~K is the temperature, $V=\frac{\pi}{4}\times50\times50\times0.9~\text{nm}^3$ is the volume of the dot, $\Delta t=1$~ps is the simulation time step and $\zeta_{\text{th}}$ is
 a Gaussian random unit vector. This equation is solved numerically and 100 write/read procedures are repeated for each value of the current/field pair applied in accordance with the experimental measurements.

$M_S$ is obtained by Vibrating Sample Magnetometer measurements, neglecting the effect of any dead layer. $V$, $T$, and $M_S$ are imposed. $\alpha$ is initially taken to be 0.033 \cite{soucaille_probing_2016, vigier2023} and $\theta_{\text{SH}}$  and  H$_{anis}^{eff}$ are first roughly estimated by fitting the SOT current vs H$_{\text{X}}$ forward switching curve using the equation established in \cite{lee_threshold_2013}:

\begin{equation}
J_{\text{SOT}} = \frac{2e}{\hbar} \frac{M_S t_m}{\theta_{\text{SH}}} 
\left( \frac{H_{\text{anis}}^{\text{eff}}}{2} - \frac{H_{\text{X}}}{\sqrt{2}} \right)
\end{equation}

$\theta_{\text{SH}}$, H$_{anis}^{eff}$, $\alpha$ and $\beta$ are then adjusted more finely to obtain the best agreement between experimental results and simulations. This is obtained with the values of the various parameters reported in Table~\ref{tab:simulation_parameters}.

\begin{table}[b]
\caption{\label{tab:simulation_parameters}
Values of the numerical parameters used for the macrospin simulations. $M_S$ is obtained by VSM measurement neglecting any dead layer, tm is controlled by the deposition, V, T and $\Delta$t are imposed. $\alpha$, $\theta_{\text{SH}}$, $\beta$ and H$_{anis}^{eff}$ are obtained by fitting simulation results to experimental results.
}
\begin{ruledtabular}
\begin{tabular}{cc}
\textrm{Parameter}&\textrm{Value}\\
\colrule
Saturation Magnetization, $M_S$&$10^6$~A/m\\
Effective Anisotropy, H$_{anis}^{eff}$& $4413$~Oe\\
Magnetic layer thickness, $t_M$& $0.9$~nm\\
Magnetic damping parameter, $\alpha$&$0.035$\\
Temperature, T&$300$~K\\
Spin Hall angle, $\theta_{\text{SH}}$&$-0.3385$\\
FL-to-DL ratio, $\beta$&$0.122$\\
Volume, V&$\frac{\pi}{4}\times(50)^2\times 0.9$~nm$^3$\\
Simulation time step,$\Delta$t&1~ps\\
Pulse Width, t$_p$& 5~ns\\
Rise/Fall Time& 70~ps\\
\end{tabular}
\end{ruledtabular}
\end{table}

The comparison of the results of the numerical simulations and those of the experiments is shown in Fig.~\ref{fig:wer_colormaps}~(d), (e) and (f). The simulations do not reproduce all the characteristics of the experimental curves, for example, the variation of the switching current as a function of H$_{\text{Y}}$ field, which is very abrupt in the simulations compared with the behavior obtained experimentally. Nevertheless, the general agreement is very good, and the different regions of no-switching, deterministic switching with minimal WER, and loss of determinism in the switching, are reproduced very well. Furthermore, the parameter values used to obtain this agreement are realistic and very close to those reported in the literature for W/CoFeB/MgO multilayers (see for example \cite{shao_roadmap_2021} and references herein). This is particularly noteworthy given that, as explained above, the dynamics of magnetization reversal by SOT can only be described in detail using a micromagnetic model.  

This agreement between experimental and simulated results could point to the predominant role of nucleation in determining threshold currents for the boundaries between no-switching/deterministic switching and deterministic switching/back-switching for sub-$100$~nm SOT-MRAM devices \cite{krizakova_tailoring_2022}.

\begin{figure*}
\includegraphics[width=0.85\linewidth]{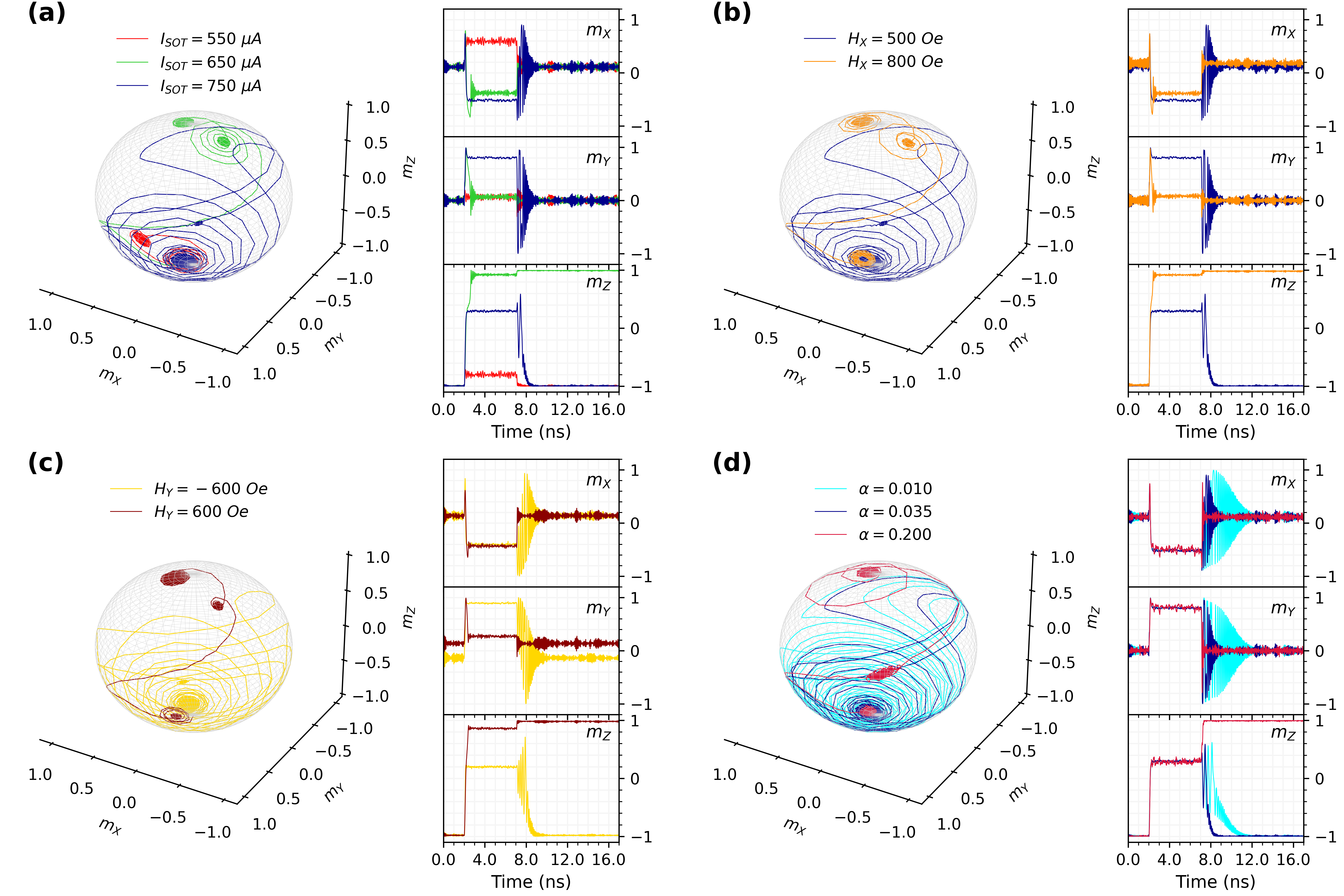}
\caption{\label{fig:switching_orbits}Magnetization switching orbits and temporal evolution of the magnetization components obtained from macrospin simulations at (a)~H$_{\text{X}}=500$~Oe for different I$_{\text{SOT}}=550~\mu$A (sub-critical)(red), $650~\mu$A (deterministic)(green) and $750~\mu$A (non-deterministic)(blue); (b)~I$_{\text{SOT}}=750~\mu$A for H$_{\text{X}}=500$~Oe (non-deterministic)(blue) and H$_{\text{X}}=800$~Oe (deterministic)(orange); (c)~H$_{\text{X}}=600$~Oe, I$_{\text{SOT}}=750~\mu$A in presence of an in-plane field along the y axis, H$_{\text{Y}}=-600$~Oe (non-deterministic) (yellow) and H$_{\text{Y}}=600$~Oe (deterministic) (brown); (d)~H$_{\text{X}}=500$~Oe, I$_{\text{SOT}}=750~\mu$A for different $\alpha=0.01$ (non-deterministic)(light blue), $0.035$ (non-deterministic)(blue) and $0.2$ (deterministic)(red).}
\end{figure*}

We can use these simulations to better understand the BSW mechanism. The study of calculated reversal orbits points to the important role of the magnetization direction at the end of the pulse, and therefore at the start of the relaxation process, on the final magnetization state. These simulations also enable us to understand the role of various parameters, such as the ratio of FL and DL terms or the magnetic damping, on the critical reversal current and the BSW threshold current, through their influence on this position of dynamic magnetization equilibrium. 

In the presence of an applied in-plane field, H$_{\text{X}}$, the two equilibrium positions m$_{\text{Z}} = \pm1$ are shifted towards the x-direction such that their new coordinates can be given by m$_+ = (\sin\theta, 0, \cos\theta)$ and m$_- = (\sin\theta, 0, -\cos\theta)$ where $\theta = \sin^{-1}(|H_{\text{X}}|/H_{anis}^{eff})$, as explained in \cite{taniguchi_theoretical_2019}. In our simulations for I$_{\text{SOT}}> 0$ and H$_{\text{X}}> 0$, the initial state of magnetization is m$_-$. During the application of the current pulse, the magnetization reaches a state of dynamic equilibrium under the predominant action of SOT, the applied magnetic field and the anisotropy. At the end of the pulse, the magnetization relaxes under the effect of the applied magnetic field, the anisotropy and thermal fluctuations. 

In the subcritical region, the current-induced torques displace the magnetization from its initial state m$_-$ but it remains close to m$_-$ and far from an in-plane position. During relaxation, the magnetization returns to its original position, as shown by I$_{\text{SOT}} = 550~\mu$A and H$_{\text{X}} = 500$~Oe in Fig.~\ref{fig:switching_orbits}~(a) (red curve).

When the current is increased, in the region of deterministic magnetization reversal, the current-induced torques are strong enough to drive the magnetization to a dynamic equilibrium close to m$_+$ with small negative x and positive y components. At the end of the current pulse, the magnetization then relaxes to the m$_+$ state, resulting in a deterministic switching, as shown in Fig.~\ref{fig:switching_orbits}~(a) for I$_{\text{SOT}} = 650~\mu$A and H$_{\text{X}} = 500$~Oe (green curve). By further increasing the current, the dynamic equilibrium begins to shift towards a fully in-plane position. This position is energetically unstable, and the resulting final state after the relaxation is highly sensitive to any energy or time fluctuation, as reported in \cite{taniguchi_theoretical_2019}. As shown in Fig.~\ref{fig:switching_orbits}~(a) (blue curve) for I$_{\text{SOT}} = 750~\mu$A and H$_{\text{X}} = 500$~Oe, the magnetization can then return to its original m$_-$ state, resulting in a BSW phenomenon.

The amplitude of the applied magnetic field influences the triggering of this BSW phenomenon.  Fig.~\ref{fig:switching_orbits}~(b) compares the switching orbits for I$_{\text{SOT}}= 750~\mu$A, H$_{\text{X}} = 500$~Oe (blue curve) and I$_{\text{SOT}} = 750~\mu$A, H$_{\text{X}} = 800$~Oe (orange curve) and shows a deterministic switching at H$_{\text{X}} = 800$~Oe at the same I$_{\text{SOT}}$. At higher applied magnetic fields H$_{\text{X}}$, the state of dynamic equilibrium departs from an in-plane position, and higher current-induced torques are required to bring it back. This state then moves out of the unstable energy region, and the reversal is deterministic. This explains the increase of the region of deterministic magnetization reversal at higher H$_{\text{X}}$ for the same I$_{\text{SOT}}$, as shown in Fig.~\ref{fig:wer_colormaps}~(d).

With the same arguments, we can infer the impact of H$_{\text{Y}}$, and consequently H$_{\text{FL}}$, by comparing the switching orbits for I$_{\text{SOT}} = 750~\mu$A and H$_{\text{X}} = 600$~Oe in the presence of a magnetic field H$_{\text{Y}}=+600$~Oe and $-600$~Oe as shown in Fig.~\ref{fig:switching_orbits}~(c). By expanding equation~\ref{eqn:llg}, we obtained the torque components due to the current-induced SOT ($\tau_\mathrm{SOT}$) and externally applied H$_\mathrm{Y}$ ($\tau_{\mathrm{H}_\mathrm{Y}}$). The z-component of $\tau_\mathrm{SOT} + \tau_{\mathrm{H}_\mathrm{Y}}$ is given by equation~\ref{eqn:zcomp_torq}.

\begin{equation}\label{eqn:zcomp_torq}
\begin{aligned}
    \tau_\mathrm{SOT,Z} + \tau_{\mathrm{H}_\mathrm{Y},Z} &= m_\mathrm{X} \Big((\beta-\alpha)\mathrm{H}_\mathrm{DL}-\mathrm{H}_\mathrm{Y} \Big) \\
    &+ \Big((1+\alpha\beta)\mathrm{H}_\mathrm{DL}-\alpha\mathrm{H}_\mathrm{Y} \Big)m_\mathrm{Y}m_\mathrm{Z}
\end{aligned}
\end{equation}

With the parameter values given in Table~\ref{tab:simulation_parameters}, the second term in equation~\ref{eqn:zcomp_torq} is dominated by the current-induced term, $(1+\alpha\beta)\mathrm{H}_\mathrm{DL}$, with negligible effect of $\mathrm{H}_\mathrm{Y}$. Whereas in the first term in equation~\ref{eqn:zcomp_torq}, $(\beta-\alpha)\mathrm{H}_\mathrm{DL}$ and $\mathrm{H}_\mathrm{Y}$ are comparable and compete. Under a positive injected current both $\mathrm{H}_\mathrm{DL}$ and m$_\mathrm{X}$ are negative resulting in overall positive out-of-plane torques pulling the magnetization towards m$_\mathrm{Z}=+1$. Starting from H$_{\text{Y}}=0$ (Fig.~\ref{fig:switching_orbits}~(a) blue curves), the application of H$_{\text{Y}}<0$ (H$_{\text{Y}}=-600$~Oe) opposes the z-component of the current-induced torques, reducing the out-of-plane torques. This reduction in the z-component of the torques further shifts the dynamic equilibrium in-plane, reinforcing the non-deterministic character of the reversal, as shown by the yellow curve in Fig.~\ref{fig:switching_orbits}~(c). On the other hand, increasing the z-component of the torques pulls the dynamic equilibrium away from this energetically unstable position. Consequently, the application of H$_{\text{Y}}=600$~Oe leads to a deterministic reversal, as shown by the brown curve in Fig.~\ref{fig:switching_orbits}~(c). Further details are provided in \cite{supplementary}. This discussion also provides insights into the impact of H$_{\text{FL}}$ ($=\beta \text{H}_\text{DL}$) since it is modulated by H$_{\text{Y}}$. It should be noted that these results are in line with previous studies showing deterministic switching for $\frac{\text{H}_{\text{FL}}}{\text{H}_{\text{DL}}}>0$ and non-deterministic switching for $\frac{\text{H}_{\text{FL}}}{\text{H}_{\text{DL}}}<0$ \cite{lee_oscillatory_2018, yoon_anomalous_2017, krizakova_tailoring_2022}. In the second case, the two terms in equation~\ref{eqn:zcomp_torq} are opposite in sign and compete to further reduce the out-of-plane torque component, contributing to shifting the dynamic equilibrium towards an in-plane position. Although previous studies \cite{lee_oscillatory_2018, yoon_anomalous_2017, krizakova_tailoring_2022} considered the competition between H$_{\text{FL}}$ and H$_{\text{DL}}$ for switching determinism, our discussion above highlights the competition between H$_{\text{FL}}-\alpha\text{H}_{\text{FL}}$ (or $(\beta-\alpha)\text{H}_\text{DL}$) and H$_{\text{DL}}$. Since $\beta \gg \alpha$ in Ta-based devices \cite{lee_oscillatory_2018, yoon_anomalous_2017, krizakova_tailoring_2022}, neglecting the $\alpha\text{H}_\text{DL}$ term is reasonable and results in identical discussions.

Since the final state depends strongly on the relaxation process, it should also depend on the magnetic damping parameter $\alpha$ as reported in \cite{lee_threshold_2013, park_macrospin_2014, taniguchi_reduction_2020}. Fig.~\ref{fig:switching_orbits}~(d) shows the switching orbits for I$_{\text{SOT}} = 750~\mu$A, H$_{\text{X}} = 500$~Oe and different $\alpha$. The damping coefficient $\alpha=0.035$, which we used in our simulations, results in slow energy dissipation and hence randomness in the final state (blue curve). This randomness is even more prevalent when $\alpha$ is lower, as shown by the light blue curve for $\alpha=0.01$. As $\alpha$ increases, energy is dissipated more rapidly and randomness is reduced, eventually disappearing, as shown by the red curve for $\alpha=0.2$ and as reported previously \cite{taniguchi_reduction_2020}.  Unlike STT switching current, which scales with $\alpha$, SOT switching current is little affected by a large magnetic damping parameter \cite{lee_threshold_2013, taniguchi_critical_2015}. Increasing the magnetic damping parameter can therefore significantly reduce the BSW without severely affecting the SOT-MRAM write current. Finally, since thermal fluctuations introduce randomness in the current-induced torques during the relaxation process, higher temperatures show an increase in WER. At low temperature, the thermal fluctuations are minimal and the final state depends only on the dynamic equilibrium and $\alpha$.

\section{Reducing the backswitching by shaping the current pulse}

\begin{figure}
\includegraphics[width=\linewidth]{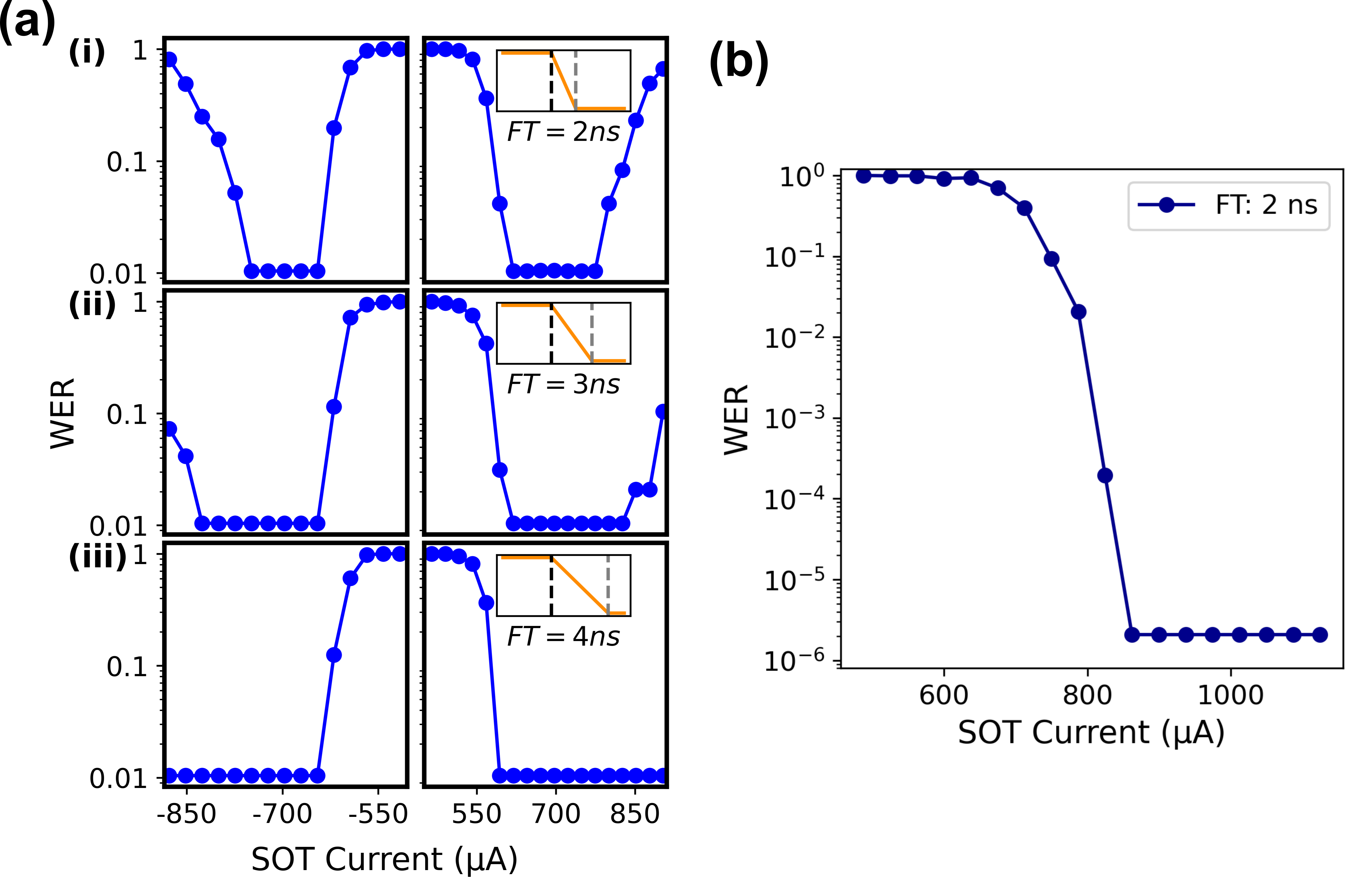}
\caption{\label{fig:fall_time}Experimental WER curves as a function of applied SOT current using a 10ns long write pulse (a) for a 100~nm diameter CoFeB~(0.9~nm) pillars at the center of $\beta$-W Hall crosses, annealed at 410\textdegree~C for 30~minutes, with fall times of (i)~2~ns, (ii)~3~ns and (iii)~4~ns and an in-plane field of 950~Oe; and (b) for a 75~nm [CoFeB~(0.72~nm)/MgO~(RA$\sim$20~$\Omega\mu$m$^2$)/Pinned Layer] dot sitting on a 130~nm wide W(4~nm) SOT track, annealed at 410\textdegree~C for 30~minutes, for a fall time of 2~ns and an in-plane field of 800~Oe.}
\end{figure}

There is another way of reducing BSW and improving WER without having to change the material system. It consists of increasing the length of the fall time of the write current pulse, as predicted numerically \cite{zhu_threshold_2020}. Fig.~\ref{fig:fall_time}~(a) shows WER curves obtained experimentally in the presence of an in-plane field, H$_X=950$~Oe and injecting 10~ns-long SOT current pulses. A widening of the forward switching region is observed as the fall time increases from 2~ns to 4~ns, with the BSW phenomenon disappearing for the current range used in the case of a fall time of 4~ns. This can be understood using the same arguments as above, by looking at magnetization relaxation at the end of the pulse. When the fall time is non-zero, current-induced torques are present during relaxation. The longer the fall time, the longer their amplitude remains at high values. The impact of thermal fluctuations is then reduced, but more importantly, the magnetization is gradually brought out of an energetically unstable position. In this way, the dynamic equilibrium is gradually modified. Such behavior is in line with the explanation given in \cite{zhu_threshold_2020}. The effect of this increase in fall time duration in the presence of a transverse field H$_\text{Y}$ or an out-of-plane field H$_\text{Z}$ has not been experimentally tested, but a discussion based on macrospin simulations can be found in \cite{supplementary}. However, much longer fall times (4~ns) were required in our experiments compared with those obtained in the simulations of \cite{zhu_threshold_2020}, typically of the order of 0.2~ns. It may be attributed to thermal fluctuations and Joule heat effects, which are not taken into account in the simulations. This difference needs to be investigated. In any case, the results of our experiments show a simple solution to the BSW phenomenon that can be easily implemented without the need for further material developments. We tested this solution on a complete SOT-MRAM single cell (Fig.~\ref{fig:fall_time}~(b)). We improved the WER by almost 3 orders of magnitude using a longer fall time of 2~ns, and obtained no error for $5\times10^5$ write attempts.

\section{Conclusion}

A detailed study of the intrinsic phenomenon of BSW in SOT-MRAMs using a statistical approach has enabled us to show that this phenomenon is statistically non-random. While SOT magnetization reversal is commonly based on a nucleation-propagation mechanism, a macrospin model using measured magnetic parameters or typical of the samples studied describes the obtained experimental results very well. This model enables us to understand the mechanisms involved in this BSW phenomenon, and in particular the role of magnetization direction at the end of the current pulse. This model also enabled us to find a solution to reduce this BSW phenomenon in order to extend the deterministic operating range. This solution was validated experimentally and very low write error rates were obtained for an extended range of current values.

\nocite{associated_data_2025}

\begin{acknowledgments}
We thank the staff of Antaios for their support throughout this project. We thank Chando Park, Minrui Yu, Luc Thomas and Mahendra Pakala from Applied Materials for fruitful discussions. This project has received funding from the European Union’s Horizon 2020 research and innovation programme under the Marie Skłodowska-Curie Grant Agreement No.~955671, and was supported by the Région Auvergne-Rhône-Alpes Pack Ambition Recherche Program (grant 19-009938-01-MAPS) and the French RENATECH network implemented at the Upstream Technological Platform in Grenoble PTA (ANR-22-PEEL-0015).
\end{acknowledgments}

% Create the reference section using BibTeX:
\bibliography{backswitching}

%%% Supplementary material %%%
\onecolumngrid
\newpage
\begin{center}
\section*{Supplementary material} 
\vspace{0.5cm}
\textbf{\large Intrinsic back-switching phenomenon in SOT-MRAM devices}
\text{Kuldeep Ray,$^{1, 2}$ Jérémie Vigier,$^{1, 2}$ Perrine Usé,$^{2}$ Sylvain Martin,$^{2}$} \text{Nicolas Lefoulon,$^{2}$ Chloé Bouard,$^{2}$ Marc Drouard,$^{2}$ and Gilles Gaudin$^{1}$}\\
\vspace{0.5cm}
\textit{$^{1}$ Univ. Grenoble Alpes, CEA, CNRS, Grenoble-INP, SPINTEC, 38000 Grenoble, France} \\
\textit{$^{2}$ Antaios, 38240 Meylan, France}
\end{center}

% reset figure labels
\renewcommand{\thefigure}{S\arabic{figure}}
\setcounter{figure}{0}

% set section counting to Note X
\renewcommand\thesection{S\arabic{section}}
\setcounter{section}{0}

\renewcommand{\theequation}{S\arabic{equation}}
\setcounter{equation}{0}

\section{Magnetic Characterization}

\subsection{Hysteresis curve} \label{hyst}
The stack $\beta$-W(4)/CoFeB(0.9)/MgO(RA$\sim$20)/CoFeB(0.3)/Ta(1)/Ru(5), where thicknesses are expressed in nanometers, was deposited via sputtering and annealed at 350\textdegree C for 30~minutes. This stack was used for all measurements in this article, except for those presented in Figure 5.

\begin{figure}[ht]
\includegraphics[width=0.85\linewidth]{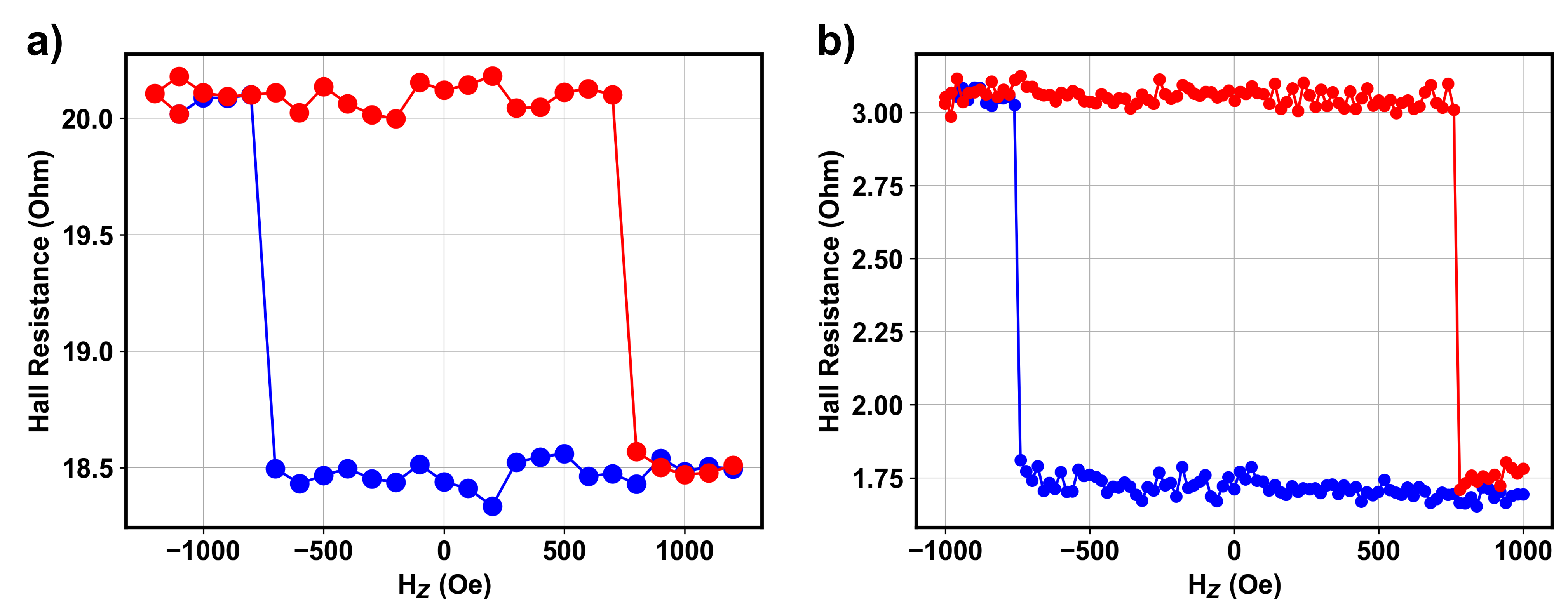}
\caption{\label{fig:s1} Hysteresis curve measured on: a) a 60~nm-diameter CoFeB(0.9)/MgO(RA$\sim$20)/CoFeB(0.3)/Ta(1)/Ru(5) pillar positioned on a 100~nm-wide $\beta$-W(4) Hall cross, annealed at 350\textdegree C for 30~minutes, and b) a 100~nm-diameter CoFeB(0.9)/MgO (RA$\sim$20)/CoFeB(0.3)/Ta(1)/Ru(5) pillar positioned on a 200~nm-wide $\beta$-W(4) Hall cross, annealed at 410\textdegree C for 30~minutes.}
\end{figure}

The stack was patterned into a 60~nm-diameter magnetic pillar positioned on a 100~nm-wide Hall cross. The Hall signal was measured by injecting a 10~$\mu$A current into the longitudinal arm, while the voltage was recorded across the transverse arm as an out-of-plane magnetic field was swept. The resulting hysteresis curve, shown in Figure~\ref{fig:s1}~a), exhibits a square shape, confirming perpendicular magnetic anisotropy, with a coercive field of approximately 750~Oe.

The results presented in Figure 5 of the main text were obtained from the same stack, but annealed at 410\textdegree C. These variations in annealing temperature were not intentionally chosen to illustrate a specific effect but resulted from an initial optimization study and sample availability. In this case, the magnetic pillar has a diameter of 100~nm, while being placed on a 200~nm-wide $\beta$-W track. The corresponding hysteresis curve, shown in Figure~\ref{fig:s1}~b), is very similar to that in Figure~\ref{fig:s1}~a), again exhibiting a coercive field of approximately 750 Oe.

\subsection{Anisotropy and thermal stability factor}

The anisotropy field $H_k$ and the thermal stability factor $\Delta$ were measured on the same Hall crosses using the switching field distribution (SFD) method. The probability of magnetization reversal was determined by continuously sweeping the applied magnetic field. The measured switching probability $P_{SW}$ is related to $H_k$ and $\Delta$ by [28, 29]:

\begin{equation}
    P_{SW} = 1 - \exp\left(-\frac{H_kf_0\sqrt{\pi}}{2r\sqrt{\Delta}}\text{erfc}\left(\sqrt{\Delta}\left(1-\frac{\mid H-H_{offset}\mid}{H_k}\right)\right)\right)
\end{equation}

where $H$ is the applied magnetic field, $H_{offset}$ is the offset field extracted from the data, $f_0$ is the attempt frequency (typically 1~GHz), and $r$ is the field sweep rate, set at 594~Oe/s. For each device, up to 50 resistance $R$ vs magnetic field $H$ cycles were measured.

\begin{figure}
\includegraphics{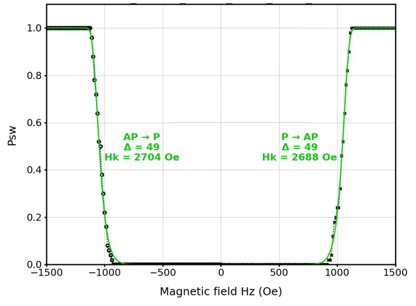}
\caption{\label{fig:s2} Example of switching probability measurements as a function of the applied magnetic field. The results are fitted using the equation in \ref{hyst} to obtain $H_k$ and $\Delta$ for both transitions: from the anti-parallel (AP) state to the parallel (P) state, and vice versa, from the P state to the AP state.}
\end{figure}

\begin{figure}
\includegraphics[width=\linewidth]{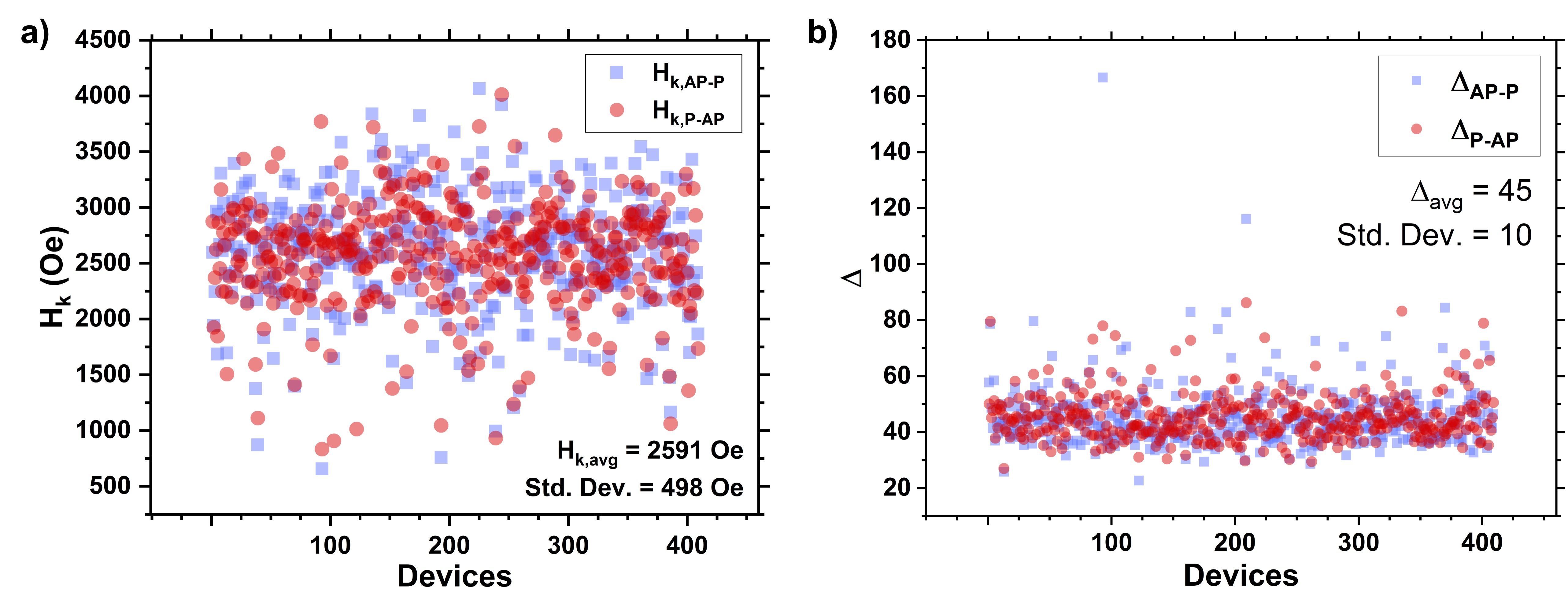}
\caption{\label{fig:s3} Distribution over 400 different devices of a) anisotropy fields $H_k$ and b) thermal stability factors $\Delta$ obtained for each of the two transitions AP-P and P-AP.}
\end{figure}

Figure~\ref{fig:s2} presents an example of switching probability measurements as a function of the applied magnetic field. The results are fitted using the equation above to extract $H_k$ and $\Delta$ for both transitions: from the anti-parallel (AP) state to the parallel (P) state, and vice versa, from the P state to the AP state. This procedure was repeated on 400 different samples. Figure~\ref{fig:s3} shows the distribution of anisotropy fields and thermal stability factors obtained for each of the two transitions. A fairly broad distribution of these parameters can be observed for each sample. The mean values obtained are respectively $H_k= 2591~$Oe (standard deviation = 498) and $\Delta = 45$ (standard deviation = 10).

\newpage
\section{Development of the LLG equation and calculation of torque components}
{\color{black}
Starting from equation 1 in the main text:
\begin{equation}
        \frac{d\mathbf{m}}{dt} = -\gamma \mu_0 \left( \mathbf{m} \times \mathbf{H}_{\text{eff}} \right) + \alpha \left( \mathbf{m} \times \frac{d\mathbf{m}}{dt} \right) + \gamma \mu_0 \left( \mathbf{m} \times \mathbf{H}_{\text{FL}} \right) + \gamma \mu_0 \left( \mathbf{m} \times \mathbf{H}_{\text{DL}} \right)
\end{equation}
The damping-like and field-like effective fields ($\mathbf{H}_{\text{DL}}$ and $\mathbf{H}_{\text{FL}}$ respectively) are given by:
\begin{align}
    \mathbf{H}_{\text{DL}} &= \mathbf{m} \times H_{\text{DL}}{\mathbf{u_y}} \\
    \mathbf{H}_{\text{FL}} &= \beta H_{\text{DL}}{\mathbf{u_y}}
\end{align}

We expand the LLG equation to obtain:
\begin{align}
    \frac{d\mathbf{m}}{dt} &= -\gamma\mu_0 \left( \mathbf{m} \times \mathbf{H}_\mathrm{eff} \right) 
    + \alpha \left( \mathbf{m} \times \frac{d\mathbf{m}}{dt} \right)
    + \gamma\mu_0 \beta H_{\text{DL}} \left( \mathbf{m} \times  {\mathbf{u_y}} \right)
    + \gamma\mu_0 H_{\text{DL}} \left( \mathbf{m} \times \mathbf{m} \times  {\mathbf{u_y}} \right) \\
    \frac{d\mathbf{m}}{dt} &= -\gamma\mu_0 \left( \mathbf{m} \times \mathbf{H}_\mathrm{eff} \right)
    + \alpha \Big[ -\gamma\mu_0 \left( \mathbf{m} \times \mathbf{m} \times \mathbf{H}_\mathrm{eff} \right) 
    + \alpha \left( \mathbf{m} \times \mathbf{m} \times \frac{d\mathbf{m}}{dt} \right) + \gamma\mu_0 \beta H_{\text{DL}} \left( \mathbf{m} \times \mathbf{m} \times  {\mathbf{u_y}} \right) \nonumber \\
    &\quad + \gamma\mu_0 H_{\text{DL}} \left( \mathbf{m} \times \mathbf{m} \times \mathbf{m} \times  {\mathbf{u_y}} \right) \Big] + \gamma\mu_0 \beta H_{\text{DL}} \left( \mathbf{m} \times  {\mathbf{u_y}} \right) 
    + \gamma\mu_0 H_{\text{DL}} \left( \mathbf{m} \times \mathbf{m} \times  {\mathbf{u_y}} \right)  \\
    (1+\alpha^2) \frac{d\mathbf{m}}{dt} &= -\gamma\mu_0 \left( \mathbf{m} \times \mathbf{H}_\mathrm{eff} \right) 
    - \alpha\gamma\mu_0 \left( \mathbf{m} \times \mathbf{m} \times \mathbf{H}_\mathrm{eff} \right) 
    + \alpha\gamma\mu_0 \beta H_{\text{DL}} \left( \mathbf{m} \times \mathbf{m} \times  {\mathbf{u_y}} \right) \nonumber \\
    &\quad + \alpha\gamma\mu_0 H_{\text{DL}} \left( \mathbf{m} \times \mathbf{m} \times \mathbf{m} \times  {\mathbf{u_y}} \right) + \gamma\mu_0 \beta H_{\text{DL}} \left( \mathbf{m} \times  {\mathbf{u_y}} \right) 
    + \gamma\mu_0 H_{\text{DL}} \left( \mathbf{m} \times \mathbf{m} \times  {\mathbf{u_y}} \right) \\
    (1+\alpha^2) \frac{d\mathbf{m}}{dt} &= -\gamma\mu_0 \left( \mathbf{m} \times \mathbf{H}_\mathrm{eff} \right) 
    - \alpha\gamma\mu_0 \left( \mathbf{m} \times \mathbf{m} \times \mathbf{H}_\mathrm{eff} \right) 
    + \alpha\gamma\mu_0 \beta H_{\text{DL}} \left( \mathbf{m} \times \mathbf{m} \times  {\mathbf{u_y}} \right) \nonumber \\
    &\quad - \alpha\gamma\mu_0 H_{\text{DL}} \left( \mathbf{m} \times  {\mathbf{u_y}} \right) + \gamma\mu_0 \beta H_{\text{DL}} \left( \mathbf{m} \times  {\mathbf{u_y}} \right) 
    + \gamma\mu_0 H_{\text{DL}} \left( \mathbf{m} \times \mathbf{m} \times  {\mathbf{u_y}} \right) \\
    \label{eqn:llg_simplified}
    (1+\alpha^2) \frac{d\mathbf{m}}{dt} &= -\gamma\mu_0 \left( \mathbf{m} \times \mathbf{H}_\mathrm{eff} \right) 
    - \alpha\gamma\mu_0 \left( \mathbf{m} \times \mathbf{m} \times \mathbf{H}_\mathrm{eff} \right) 
    + \gamma\mu_0 (1+\alpha\beta) H_{\text{DL}} \left( \mathbf{m} \times \mathbf{m} \times  {\mathbf{u_y}} \right) \nonumber \\
    &\quad + \gamma\mu_0 (\beta - \alpha) H_{\text{DL}} \left( \mathbf{m} \times  {\mathbf{u_y}} \right)
\end{align}

In the presence of $H_\mathrm{X}$ and $H_\mathrm{Y}$, $\mathbf{H}_\mathrm{eff}$ can be written as $(H_\mathrm{X}, H_\mathrm{Y}, H_\mathrm{ani})$. Using this value of $\mathbf{H}_\mathrm{eff}$, equation~\ref{eqn:llg_simplified} can be separated into $x$, $y$ and $z$ components:

\begin{align}
    \label{eqn:dmdt_xcomp}
    (1+\alpha^2) \frac{dm_\mathrm{X}}{dt} &= -\gamma\mu_0 (m_\mathrm{Y} H_\mathrm{ani} - m_\mathrm{Z} H_\mathrm{Y}) 
    - \alpha\gamma\mu_0 \left( m_\mathrm{X} m_\mathrm{Y} H_\mathrm{Y} - (m_\mathrm{Y}^2 + m_\mathrm{Z}^2) H_\mathrm{X} + m_\mathrm{X} m_\mathrm{Z} H_\mathrm{ani} \right) \nonumber \\
    &\quad + \gamma\mu_0 (1+\alpha\beta) H_{\text{DL}} m_\mathrm{X} m_\mathrm{Y} 
    - \gamma\mu_0 (\beta - \alpha) H_{\text{DL}} m_\mathrm{Z} \\
    \label{eqn:dmdt_ycomp}
    (1+\alpha^2) \frac{dm_\mathrm{Y}}{dt} &= -\gamma\mu_0 (m_\mathrm{Z} H_\mathrm{X} - m_\mathrm{X} H_\mathrm{ani}) 
    - \alpha\gamma\mu_0 \left( m_\mathrm{X} m_\mathrm{Y} H_\mathrm{X} + m_\mathrm{Y} m_\mathrm{Z} H_\mathrm{ani} - (m_\mathrm{X}^2 + m_\mathrm{Z}^2) H_\mathrm{Y} \right) \nonumber \\
    &\quad - \gamma\mu_0 (1+\alpha\beta) H_{\text{DL}} (m_\mathrm{X}^2 + m_\mathrm{Z}^2) \\
    \label{eqn:dmdt_zcomp}
    (1+\alpha^2) \frac{dm_\mathrm{Z}}{dt} &= -\gamma\mu_0 (m_\mathrm{X} H_\mathrm{Y} - m_\mathrm{Y} H_\mathrm{X}) 
    - \alpha\gamma\mu_0 \left( m_\mathrm{X} m_\mathrm{Z} H_\mathrm{X} + m_\mathrm{Y} m_\mathrm{Z} H_\mathrm{Y} - (m_\mathrm{X}^2 + m_\mathrm{Y}^2) H_\mathrm{ani} \right) \nonumber \\
    &\quad + \gamma\mu_0 (1+\alpha\beta) H_{\text{DL}} m_\mathrm{Y} m_\mathrm{Z} 
    + \gamma\mu_0 (\beta - \alpha) H_{\text{DL}} m_\mathrm{X}.
\end{align}

On rearranging the different terms in the equations~\ref{eqn:dmdt_xcomp},~\ref{eqn:dmdt_ycomp} and~\ref{eqn:dmdt_zcomp}, we obtained:
\begin{align}
    \frac{(1+\alpha^2)}{\gamma\mu_0} \frac{dm_\mathrm{X}}{dt} &= -m_\mathrm{Y} H_\mathrm{ani} + \alpha (m_\mathrm{Y}^2 + m_\mathrm{Z}^2) H_\mathrm{X} - \alpha m_\mathrm{X} m_\mathrm{Z} H_\mathrm{ani} 
    + m_\mathrm{Z} \Big( H_\mathrm{Y} - (\beta - \alpha) H_{\text{DL}} \Big) \nonumber \\
    &\quad + \Big( (1+\alpha\beta) H_{\text{DL}} - \alpha H_\mathrm{Y} \Big) m_\mathrm{X} m_\mathrm{Y} \\
    \frac{(1+\alpha^2)}{\gamma\mu_0} \frac{dm_\mathrm{Y}}{dt} &= -m_\mathrm{Z} H_\mathrm{X} + m_\mathrm{X} H_\mathrm{ani} - \alpha m_\mathrm{X} m_\mathrm{Y} H_\mathrm{X} - \alpha m_\mathrm{Y} m_\mathrm{Z} H_\mathrm{ani} 
    + \Big( \alpha H_\mathrm{Y} - (1+\alpha\beta) H_{\text{DL}} \Big) (m_\mathrm{X}^2 + m_\mathrm{Z}^2) \\
    \frac{(1+\alpha^2)}{\gamma\mu_0} \frac{dm_\mathrm{Z}}{dt} &= m_\mathrm{Y} H_\mathrm{X} - \alpha m_\mathrm{X} m_\mathrm{Z} H_\mathrm{X} + \alpha (m_\mathrm{X}^2 + m_\mathrm{Y}^2) H_\mathrm{ani} 
    + m_\mathrm{X} \Big( (\beta - \alpha) H_{\text{DL}} - H_\mathrm{Y} \Big) \nonumber \\
    &\quad + \Big( (1+\alpha\beta) H_{\text{DL}} - \alpha H_\mathrm{Y} \Big) m_\mathrm{Y} m_\mathrm{Z}
\end{align}

\begin{comment}
    Since $H_{\text{DL}}\gg \alpha H_\mathrm{Y}$, $(1+\alpha\beta) H_{\text{DL}}-\alpha H_\mathrm{Y}$ is dominated by the current induced term and the effect of $H_\mathrm{Y}$ is less pronounced. On the other hand, $(\beta-\alpha) H_{\text{DL}}$ and $H_\mathrm{Y}$ act against each other and have the same order of magnitude. Hence, $(\beta-\alpha) H_{\text{DL}}-H_\mathrm{Y}$ is strongly impacted by $H_\mathrm{Y}$.
\end{comment}

Using the parameters in the main text and $H_{\text{DL}}=\theta_\mathrm{SH}\frac{\mathrm{I}_\mathrm{SOT}}{t_\mathrm{SOT} w_\mathrm{SOT}}\frac{\hbar}{2e\mu_0 M_S t_m}$, we calculated $H_{\text{DL}}=2321~$Oe for $\mathrm{I}_\mathrm{SOT}=750~\mu$A. Since $H_{\text{DL}}\gg \alpha H_\mathrm{Y}$, $(1+\alpha\beta) H_{\text{DL}}-\alpha H_\mathrm{Y}$ is dominated by the current induced term and the influence of $H_\mathrm{Y}$ is negligible. On the other hand, $(\beta-\alpha) H_{\text{DL}}$ and $H_\mathrm{Y}$ have similar magnitudes but opposite signs. Hence, $(\beta-\alpha) H_{\text{DL}}-H_\mathrm{Y}$ is strongly impacted by $H_\mathrm{Y}$. 

\begin{figure}[ht]
\includegraphics[width=0.8\linewidth]{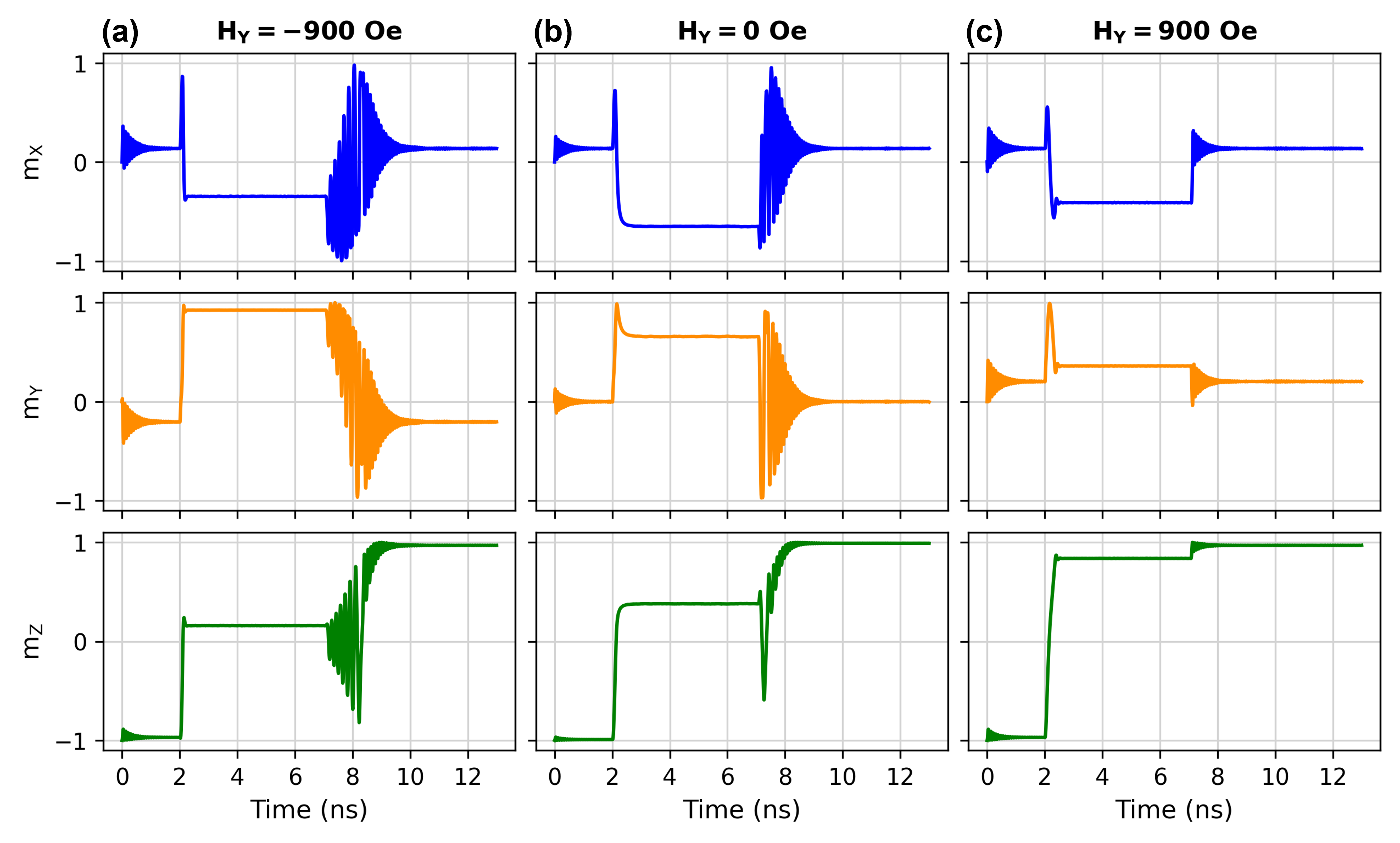}
\caption{\label{fig:s4} \textcolor{black}{Simulated temporal evolution of $m_\mathrm{X}$, $m_\mathrm{Y}$ and $m_\mathrm{Z}$ under $\mathrm{I}_\mathrm{SOT}=750~\mu$A and $H_\mathrm{X}=600~$Oe, and under an in-plane transverse field: (a)~$H_\mathrm{Y}=-900~$Oe, (b)~$H_\mathrm{Y}=0~$Oe, and (c)~$H_\mathrm{Y}=900~$Oe.}}
\end{figure}

To provide a more comprehensive discussion of the effect of $H_\mathrm{Y}$, we have added simulations for larger positive and negative $H_\mathrm{Y}$ amplitudes (see Figure~\ref{fig:s4}), completing Figure~4 in the main text. In our simulations we start from $m_\mathrm{Z}=-1$. In the absence of $H_\mathrm{Y}$~($H_\mathrm{Y}=0$), $m_\mathrm{X} (\beta-\alpha) H_{\text{DL}}$ is positive but $(1+\alpha\beta)H_{\text{DL}}m_\mathrm{Y}m_\mathrm{Z}$ is negative, since $m_\mathrm{X}$ and $H_{\text{DL}}$ are negative while $m_\mathrm{Y}$ and $m_\mathrm{Z}$ are positive. These two terms compete to pull the dynamic equilibrium position towards an in-plane position~(Figure~\ref{fig:s4}~(b)). For non-zero $H_\mathrm{Y}$ within the measured range, this second term does not change sign, as the contribution from $H_\mathrm{Y}$ is negligible and both $m_\mathrm{Y}$ and $m_\mathrm{Z}$ remain positive. However, $H_\mathrm{Y}$ does affect the first term, with its effect determined by its sign and amplitude. For $H_\mathrm{Y}>0$, $(\beta-\alpha) H_{\text{DL}}-H_\mathrm{Y}$ is negative and since $m_\mathrm{X}$ is also negative, the term $m_\mathrm{X} \Big((\beta-\alpha) H_{\text{DL}}-H_\mathrm{Y}\Big)$ pulls the magnetization towards $m_\mathrm{Z}=+1$ resulting in deterministic switching~(consistent with $H_\mathrm{Y}=600$~Oe, yellow curve in Figure~4~(c) and $H_\mathrm{Y}=900$~Oe, shown in Figure~\ref{fig:s4}~(c)). Similarly for $H_\mathrm{Y}<0$, $(\beta-\alpha) H_{\text{DL}}-H_\mathrm{Y}$ is positive if $\left\lvert H_\mathrm{Y}\right\lvert > \left\lvert(\beta-\alpha) H_{\text{DL}}\right\lvert$ and since $m_\mathrm{X}$ is negative, the term $m_\mathrm{X} \Big((\beta-\alpha) H_{\text{DL}}-H_\mathrm{Y}\Big)$ is negative and the magnetization stays in an in-plane position~(consistent with $H_\mathrm{Y}=-600$~Oe, brown curve in Figure~4~(c) and $H_\mathrm{Y}=-900$~Oe, shown in Figure~\ref{fig:s4}~(a)). Since the z-component of the torques increases with increasingly positive $H_\mathrm{Y}$, the equilibrium shifts away from the in-plane position, as shown in Figure~\ref{fig:s4}. With a more negative $H_\mathrm{Y}$, the magnetization increasingly moves towards the in-plane position~(not shown here). The subsequent relaxation from this position is precessional and leads to non-deterministic switching and higher error rates. 
}

{\color{black}

\section{Simulated WER maps for a broader range of current values and various temperatures}

To assess the influence of temperature on the WER and on both back-switching (BSW) and forward-switching (FSW) regions, we have extended the simulations from Figure~3 of the main text to two additional temperatures, 1 K and 500 K. We define the FSW region as the set of conditions under which, at room temperature, the magnetization switches from its initial direction to the opposite direction. It is typically shown in yellow in Figure~\ref{fig:s5}~(d–f). The BSW region is then defined as the set of conditions that corresponds neither to the sub-critical region (no switching observed) nor to the FSW region. We also increased the simulated current range up to $1300~\mu$A (instead of $800~\mu$A). Figure~\ref{fig:s5} presents these results for 1~K, 300~K, and 500~K under three field configurations: (a, d, g) variable H$_{\text{X}}$; (b, e, h) variable H$_{\text{Y}}$ with H$_{\text{X}}=600~$Oe; and (c, f, i) variable H$_{\text{Z}}$ with H$_{\text{X}}=800~$Oe. The static H$_{\text{X}}$ values match those used in Figure~3.

\begin{figure}[ht]
\includegraphics[width=0.9\linewidth]{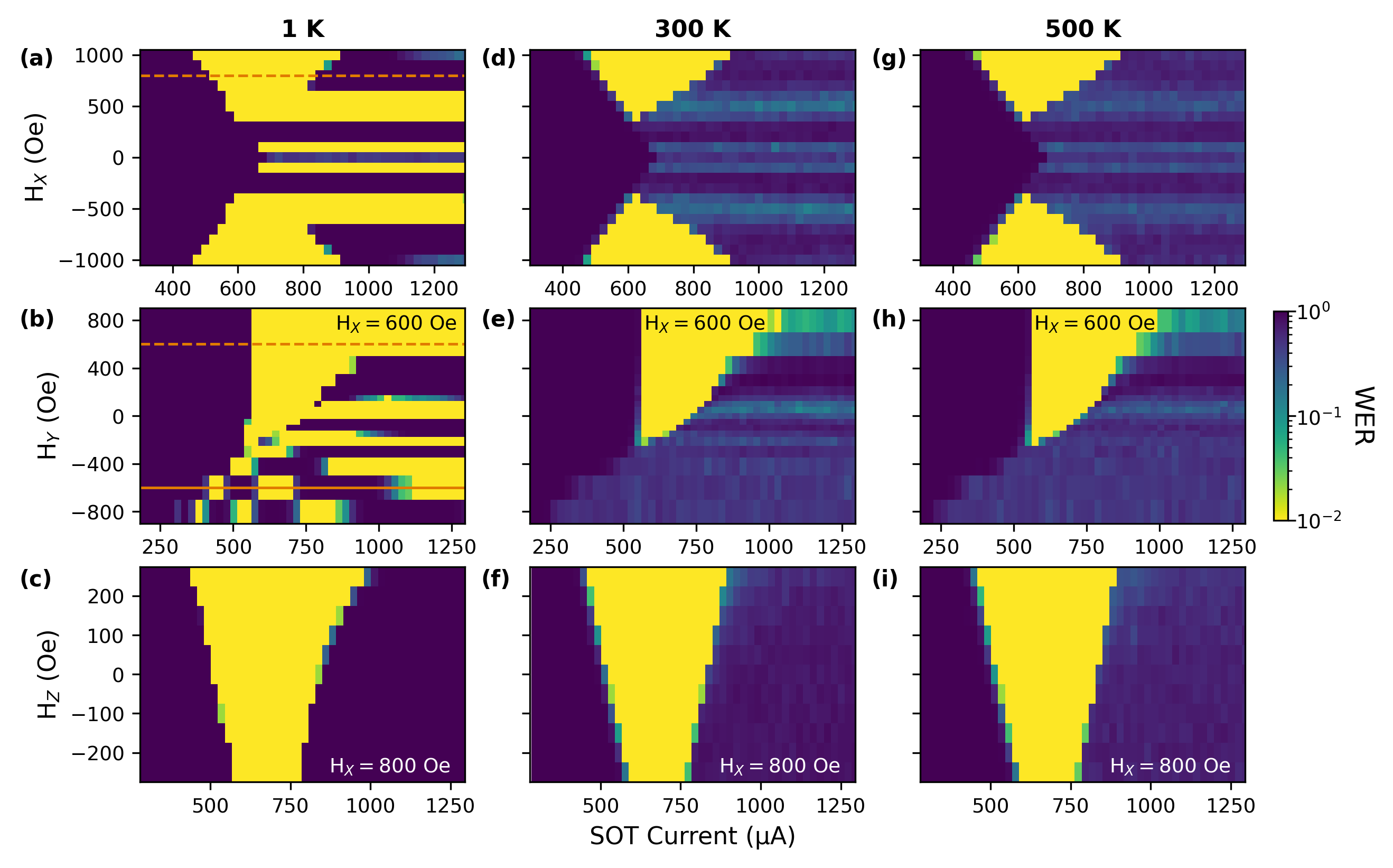}
\caption{\label{fig:s5} Simulated WER color maps as a function of applied SOT current, obtained over 100 write/read cycles, for three temperatures and three field configurations each: $\text{T}=1~$K (a) variable H$_{\text{X}}$, (b) variable H$_{\text{Y}}$ under H$_{\text{X}}=600~$Oe, (c) variable H$_{\text{Z}}$ under H$_{\text{X}}=800~$Oe; $\text{T}=300~$K (d) variable H$_{\text{X}}$, (e) variable H$_{\text{Y}}$ under H$_{\text{X}}=600~$Oe, (f) variable H$_{\text{Z}}$ under H$_{\text{X}}=800~$Oe; $\text{T}=500~$K (g) variable H$_{\text{X}}$, (h) variable H$_{\text{Y}}$ under H$_{\text{X}}=600~$Oe, (i) variable H$_{\text{Z}}$ under H$_{\text{X}}=800~$Oe. In all panels, blue ($\text{WER}=1$) denotes switching failure in every trial, while yellow indicates deterministic switching (no writing error, $\text{WER}=10^{-2}$).). The orange dashed line in (a) is drawn at H$_{\text{X}}=800~$Oe, as discussed in Figure~\ref{fig:s6}. The orange dashed (resp. continuous) line in (b) is drawn at H$_{\text{Y}}=600~$Oe (resp. $-600~$Oe) as discussed in Figure~\ref{fig:s8} (resp. \ref{fig:s9}).}
\end{figure}

At 1~K, thermal fluctuations are negligible. The final magnetization state depends only on the damping constant and the magnetization direction at the end of the current pulse, before it relaxes by precessing about the residual magnetic fields. Stochasticity is minimal, so magnetization switching is essentially deterministic: either FSW (yellow), magnetization opposite to its initial state, or no switching (blue), magnetization returns to its original orientation, or BSW (blue or yellow depending on the final orientation of the magnetization). This behavior is shown in Figure~\ref{fig:s5}~(a-c).

As temperature increases, FSW remains robust against thermal fluctuations, whereas deterministic BSW regions are strongly suppressed: deterministic BSW bands disappear, yielding zones of elevated WER, but below unity. This contrasts with the experimental behavior (cf. Figure~2 of the main text). As discussed, macrospin simulations capture the global switching behavior efficiently, enabling rapid WER mapping across parameter sweeps, but cannot accurately reproduce BSW phenomena based on domain nucleation/DW propagation processes, which would require micromagnetic modeling at prohibitive computational cost.

\subsection{H$_{\text{X}}$ dependence (Figure~\ref{fig:s5}~(a, d, g))}
At 1~K, FSW triangles appear at SOT currents in the range measured in experiments (the same triangles as the ones observed at 300 K Figure~\ref{fig:s5}~(d). Beyond this range, BSW bands appear either blue or yellow. In the absence of thermal fluctuations, these nearly horizontal bands indicate deterministic switching that alternates between “up” and “down”. 

To elucidate this pattern, we first initialized the magnetization at various orientations defined by the spherical angles $\theta$ and $\phi$, and then allowed it to relax under a static in-plane field of H$_\mathrm{X}=800~$Oe (reported in orange dashed line in Figure~\ref{fig:s5}~(a)). The results are shown in Figure~\ref{fig:s6}: the initial state is m$_\mathrm{Z}=-1$, yellow indicates switching into the opposite state, and blue indicates relaxation back to the initial state. The red dashed line marks the equatorial positions ($\theta=90$\textdegree). Under H$_\mathrm{X}=800~$Oe, the final magnetization state depends critically on its orientation at the end of the current pulse, as discussed in the main text. Next, we computed the magnetization’s end-of-pulse orientation under the same H$_\mathrm{X}=800~$Oe for injected currents ranging from $600~\mu$A to $1400~\mu$A, and overlaid these points on a zoomed-in view of the band-like pattern in Figure~\ref{fig:s6}. The resulting plot is presented in Figure~\ref{fig:s7}. We observe that lower currents ($600–800~\mu$A) drive the magnetization into the yellow FSW region at pulse termination, whereas higher currents ($900–1400~\mu$A) place it consistently within the blue BSW regions. Thus, the alternating band-like pattern seen in Figure~\ref{fig:s5}~(a) arise from two factors: (1) the current-dependent magnetization orientation at the end of the pulse, and (2) the value of H$_\mathrm{X}$ around which the magnetization precesses during relaxation.

We note that similar band-like patterns have been reported in the absence of thermal fluctuations, for example, in Figure~5(a) of \cite{zhu_threshold_2020}~(Ref. 23 in the main text) where these band-like patterns curve as H$_\mathrm{X}$ increases. This curvature is another sign of the interplay between the two factors presented above.

Raising the temperature to $300~$K preserves FSW bands, consistent with room-temperature experiments, but deterministic BSW bands vanish, leaving WER modulated by current and H$_\mathrm{X}$. Increasing the temperature to $500~$K yields no qualitative change.

\begin{figure}[ht]
\includegraphics[width=0.65\linewidth]{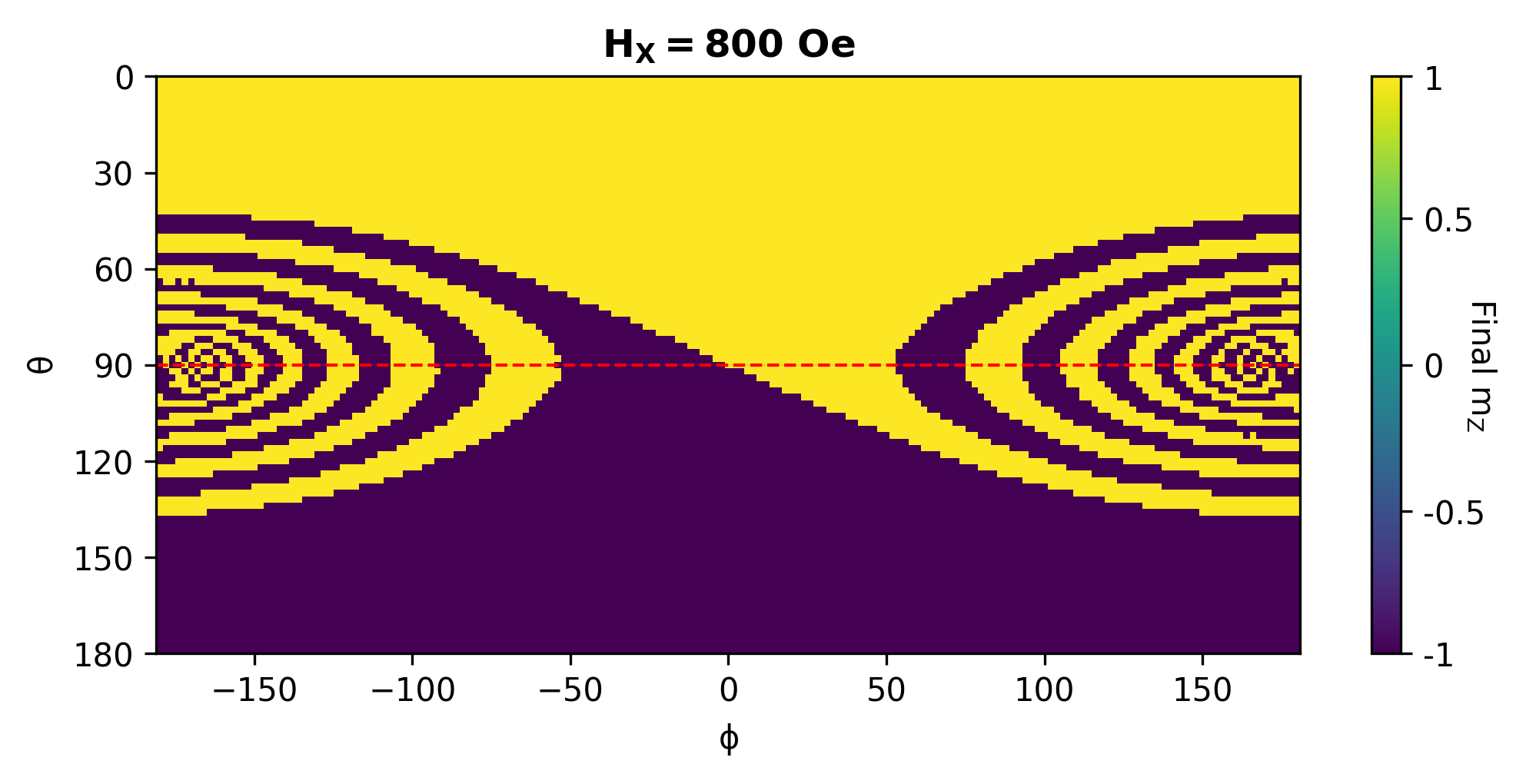}
\caption{\label{fig:s6} Simulated magnetization orientation after relaxation around an in-plane field of H$_{\text{X}}=800~$Oe, plotted as a function of the initial spherical angles $\theta$ and $\phi$. Yellow denotes a final state with m$_{\text{X}}=+1$, while blue denotes m$_{\text{X}}=-1$. These simulations were performed in the absence of thermal fluctuations. The red dashed line marks $\theta=90$\textdegree corresponding to an in-plane magnetization orientation.}
\end{figure}

\begin{figure}[ht]
\includegraphics[width=0.65\linewidth]{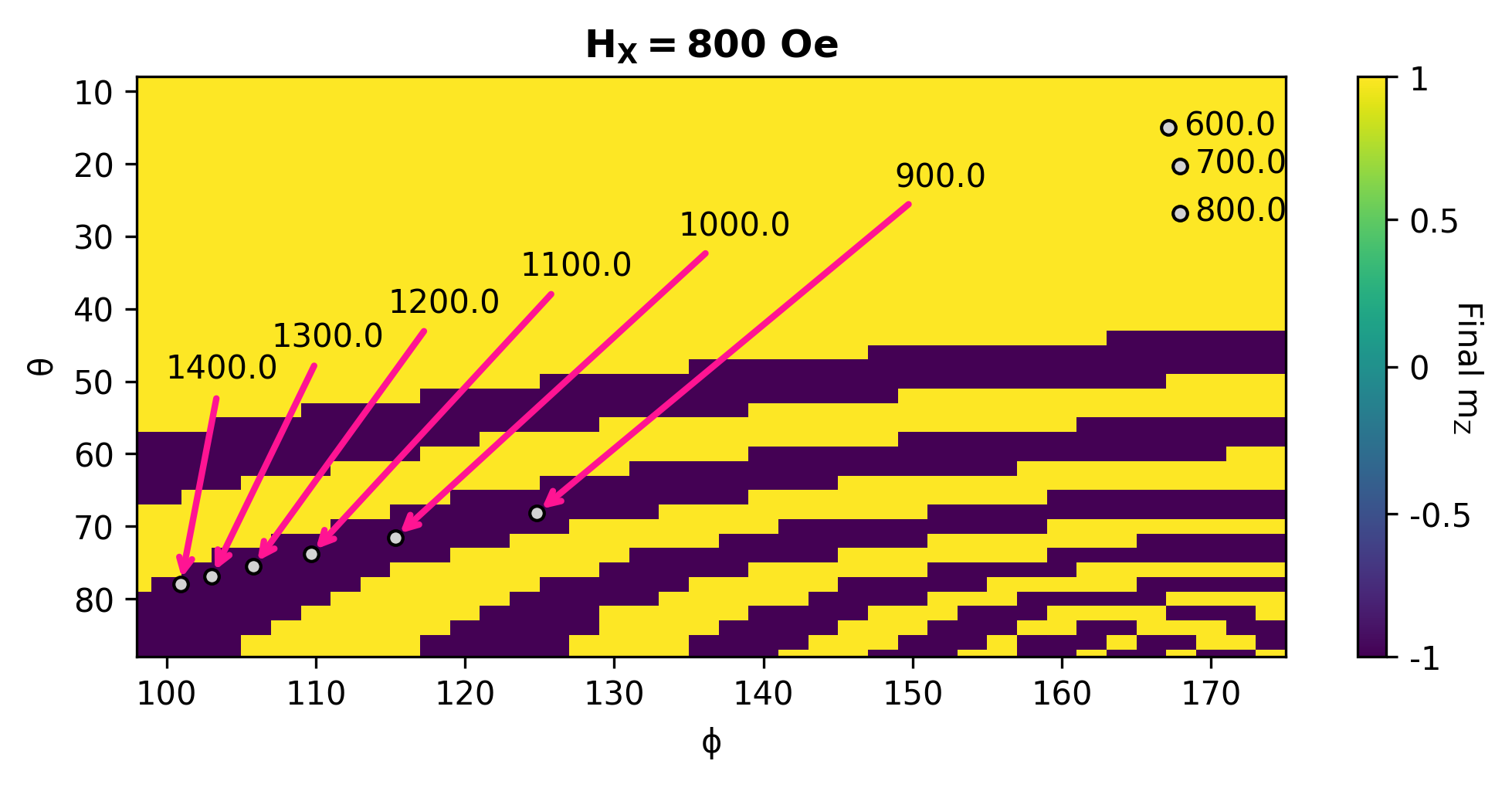}
\caption{\label{fig:s7} Zoom of Figure~\ref{fig:s6} showing the magnetization orientation at the end of the current pulse for injected currents from $600~\mu$A to $1400~\mu$A.}
\end{figure}

\subsection{H$_{\text{Y}}$ dependence under H$_{\text{X}}=600~$Oe~(Figure~\ref{fig:s5}~(b, e, h))}
At 1 K, the FSW region for H$_{\text{Y}}>0$ (as in Figure~3) is accompanied by deterministic BSW band-like patterns, analogous to the H$_\mathrm{X}$ case. When H$_\mathrm{Y}$ reverses sign, these band-like patterns curve and evolve into a checkerboard pattern (at our field and current resolution).

To understand this behavior, we repeated the previous study under fixed fields H$_\mathrm{X}=600~$Oe and H$_\mathrm{Y}=-600~$Oe, as indicated by the solid orange line in Figure~\ref{fig:s5}~(b). Figure~\ref{fig:s8}~(a) shows the result of letting the magnetization relax under these two fields, starting from orientations specified by the spherical angles $\theta$ and $\phi$. We observe that applying H$_\mathrm{Y}=-600~$Oe shifts the pattern in $\phi$ compared to Figure~\ref{fig:s6}. Next, we computed the magnetization’s end-of-pulse orientation for several currents spanning the ranges used in Figures~3 and \ref{fig:s5}. Unlike the H$_\mathrm{X}$-dependent case discussed above, the final magnetization does not follow the same color bands in Figure~\ref{fig:s8}~(a). Instead, it alternates between blue and yellow regions as current increases (Figure~\ref{fig:s8}~(b)).

\begin{figure}[ht]
\includegraphics[width=0.8\linewidth]{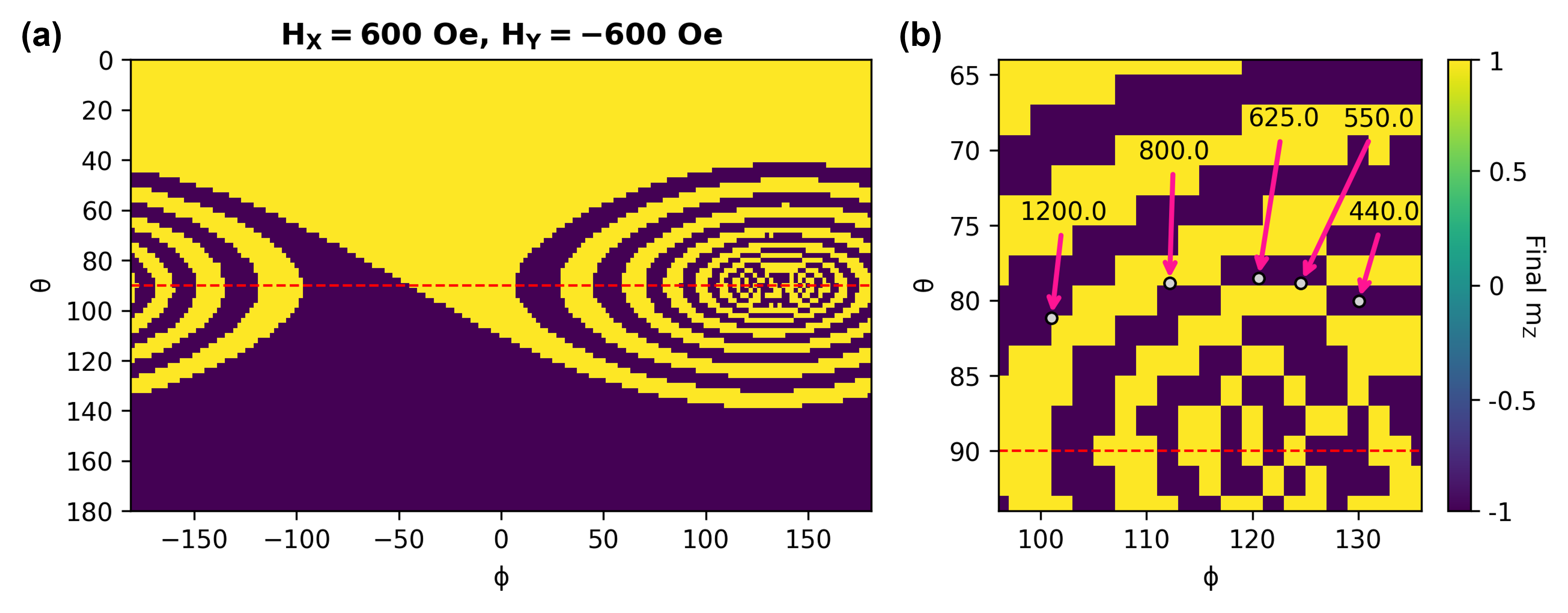}
\caption{\label{fig:s8} (a) Simulated magnetization orientation after relaxation around a combination of H$_{\text{X}}=600~$Oe and H$_{\text{Y}}=-600~$Oe, plotted as a function of the initial spherical angles $\theta$ and $\phi$. Yellow denotes a final state with m$_{\text{X}}=+1$, while blue denotes m$_{\text{X}}=-1$. These simulations were performed in the absence of thermal fluctuations. (b) Zoom of (a) showing the magnetization orientation at the end of the current pulse for injected currents from $440~\mu$A to $1200~\mu$A. The red dashed line marks $\theta=90$\textdegree corresponding to an in-plane magnetization orientation.}
\end{figure}

\begin{figure}[ht]
\includegraphics[width=0.8\linewidth]{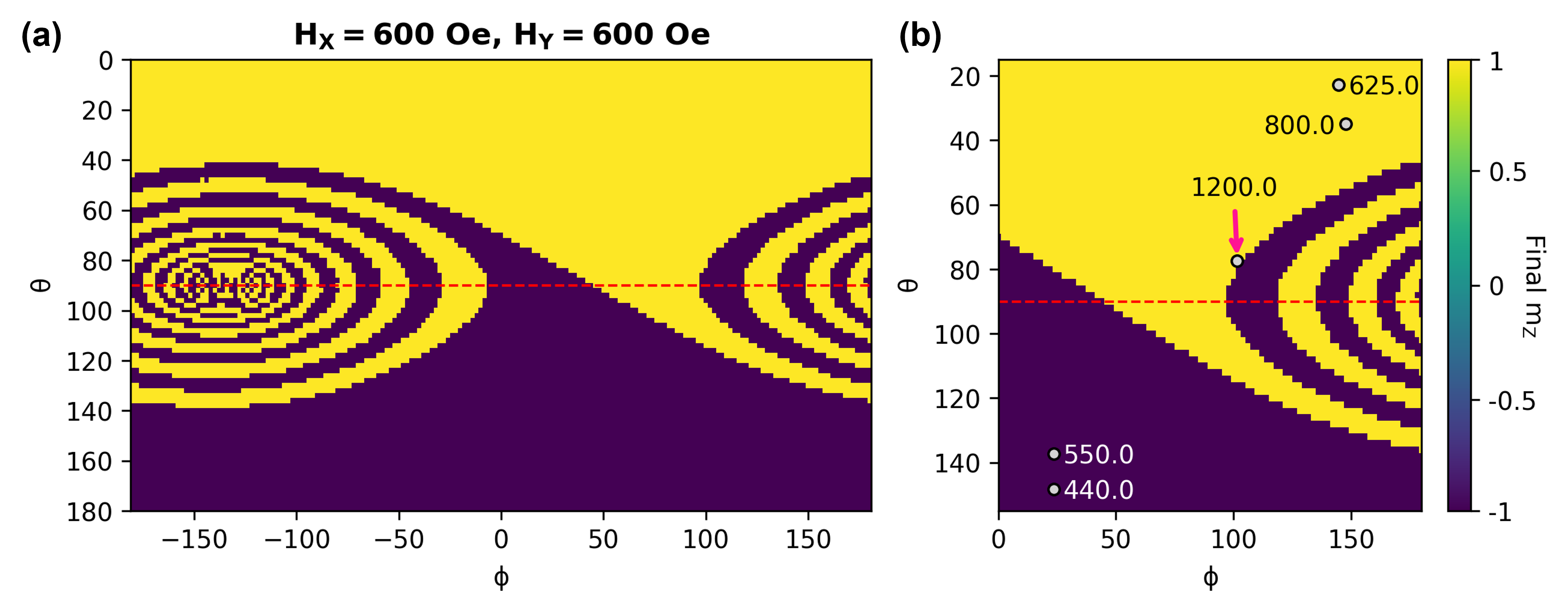}
\caption{\label{fig:s9} (a) Simulated magnetization orientation after relaxation around a combination of H$_{\text{X}}=600~$Oe and H$_{\text{Y}}=600~$Oe, plotted as a function of the initial spherical angles $\theta$ and $\phi$. Yellow denotes a final state with m$_{\text{X}}=+1$, while blue denotes m$_{\text{X}}=-1$. These simulations were performed in the absence of thermal fluctuations. (b) Zoom of (a) showing the magnetization orientation at the end of the current pulse for injected currents from $440~\mu$A to $1200~\mu$A. The red dashed line marks $\theta=90$\textdegree corresponding to an in-plane magnetization orientation.}
\end{figure}

We repeated this procedure for H$_\mathrm{Y}=600~$Oe, with results shown in Figure~\ref{fig:s9}. In Figure~\ref{fig:s9}~(a), the entire pattern is shifted towards positive $\phi$ by the presence of H$_\mathrm{Y}>0$. Figure~\ref{fig:s9}~(b) demonstrates that as current increases, the system transitions from a subcritical regime, where currents of $440~\mu$A and $550~\mu$A cannot switch the magnetization, to a switching regime for higher currents. In this regime, the magnetization at pulse end remains within the yellow band but in the FSW region for currents of $625~\mu$A and $800~\mu$A, and in the BSW band at $1200~\mu$A. Again, the band and checkerboard patterns arise from two factors: (1) the current-dependent magnetization orientation at the end of the pulse, and (2) the value of H$_\mathrm{X}$ and H$_\mathrm{Y}$ around which the magnetization precesses during relaxation.

At $300~$K, deterministic BSW bands disappear, yielding a similar overall pattern but with intermediate WER values. Notably, the FSW region vanishes as soon as H$_\mathrm{Y}$ becomes negative. This behavior can be understood by examining Figures~\ref{fig:s6}, \ref{fig:s8}, and \ref{fig:s9}. When a negative transverse field (H$_\mathrm{Y}<0$) is applied, the checkerboard pattern of blue and yellow bands shifts towards $\phi<0$ (cf. Figures~\ref{fig:s6} and \ref{fig:s8}), whereas a positive field (H$_\mathrm{Y}>0$) produces a shift towards $\phi>0$ (cf. Figure~\ref{fig:s9}). For example, under H$_\mathrm{Y}=-600~$Oe, all positions with $\phi>0$ and $\theta \approx 90$\textdegree fall within the BSW region. Moreover, as discussed in the main text and illustrated in Figure~\ref{fig:s5}~(c), a negative H$_\mathrm{Y}$ acts against the $(\beta - \alpha)H_\mathrm{DL}$ induced by a positive current reducing the z-component of the torques. These combined effects explain why deterministic switching ceases to occur once H$_\mathrm{Y}$ falls below a small negative threshold.

\subsection{H$_{\text{Z}}$ dependence under H$_{\text{X}}=800~$Oe~(Figure~\ref{fig:s5}~(c, f, i))}

At 1 K, the FSW region (yellow) coexists with deterministic BSW into the opposite direction at higher currents. In this current regime, H$_\mathrm{Z}$ is insufficient to stabilize the magnetization during post-pulse relaxation. As the applied current increases, the magnetization is driven into a progressively unstable, high-energy configuration. Its subsequent relaxation under the combined influence of H$_\mathrm{X}$ and H$_\mathrm{Z}$ then leads to BSW towards the opposite direction. Upon raising the temperature ($300~$K and $500~$K), the FSW region narrows slightly, reflecting the impact of thermal fluctuations on the end-of-pulse magnetization orientation and the subsequent magnetization relaxation.

A reference to this section has been added in the main text at the end of the discussion of Figure~4~(c).

\section{Simulations of the effect of current pulse fall time on WER under variable H$_\mathrm{X}$, H$_\mathrm{Y}$, and H$_\mathrm{Z}$ fields}

The WER color maps as a function of injected SOT current were simulated for three field configurations: variable H$_\mathrm{X}$, variable H$_\mathrm{Y}$ with H$_\mathrm{X}=600~$Oe, and variable H$_\mathrm{Z}$ with H$_\mathrm{X}=800~$Oe, using three different pulse fall times: (a–c)~$70~$ps (as in the experiments and the simulations of Figure~3), (d–f)~$500~$ps, and (g–i)~$1~$ns.

\subsection{H$_{\text{X}}$ dependence (Figure~\ref{fig:s10}~(a, d, g))}

A very strong effect of the pulse fall time on the WER is observed. Increasing the fall time extends the FSW region to much higher currents and fields, thereby greatly widening the operational window. For a fall time of 1~ns, the entire explored current range exhibits deterministic switching into the state opposite to the initial magnetization, except for a low-field stripe where H$_\mathrm{X}$ is insufficient to initiate switching. As discussed in the main text, although the simulations reproduce the qualitative trend seen experimentally, the characteristic time scales differ substantially.

\subsection{H$_{\text{Y}}$ dependence under H$_{\text{X}}=600~$Oe~(Figure~\ref{fig:s10}~(b, e, h))}

Here too, the pulse fall time has a pronounced impact on WER: extending the fall time dramatically enlarges the FSW region towards higher currents. However, this effect does not apply for negative H$_{\text{Y}}$. A possible explanation is that, during the finite fall time relaxation, the H$_{\text{Y}}<0$ field remains active and shifts the band-like pattern, as shown in Figure~\ref{fig:s8}. Consequently, the magnetization is held within an unstable region throughout its relaxation, which may explain why a finite fall time cannot improve switching under these conditions: the benefit of a gradual fall, which under a purely H$_{\text{X}}$ field would quasi-statically steer the magnetization away from the equator in a symmetric band-like pattern, is therefore absent here. This could account for the sharply defined yellow (FSW) and blue (BSW) regions near H$_{\text{Y}}=0$. 

\subsection{H$_{\text{Z}}$ dependence under H$_{\text{X}}=800~$Oe~(Figure~\ref{fig:s10}~(c, f, i))}

Again, increasing the fall time greatly expands the FSW region across the entire high-current range used in our simulations. The underlying mechanism is the same as described in the main text. Note that the range of H$_{\text{Z}}$ fields simulated here matches the experimental range and is narrower than those used for H$_{\text{X}}$ and H$_{\text{Y}}$. 

\begin{figure} [ht]
\includegraphics[width=0.9\linewidth]{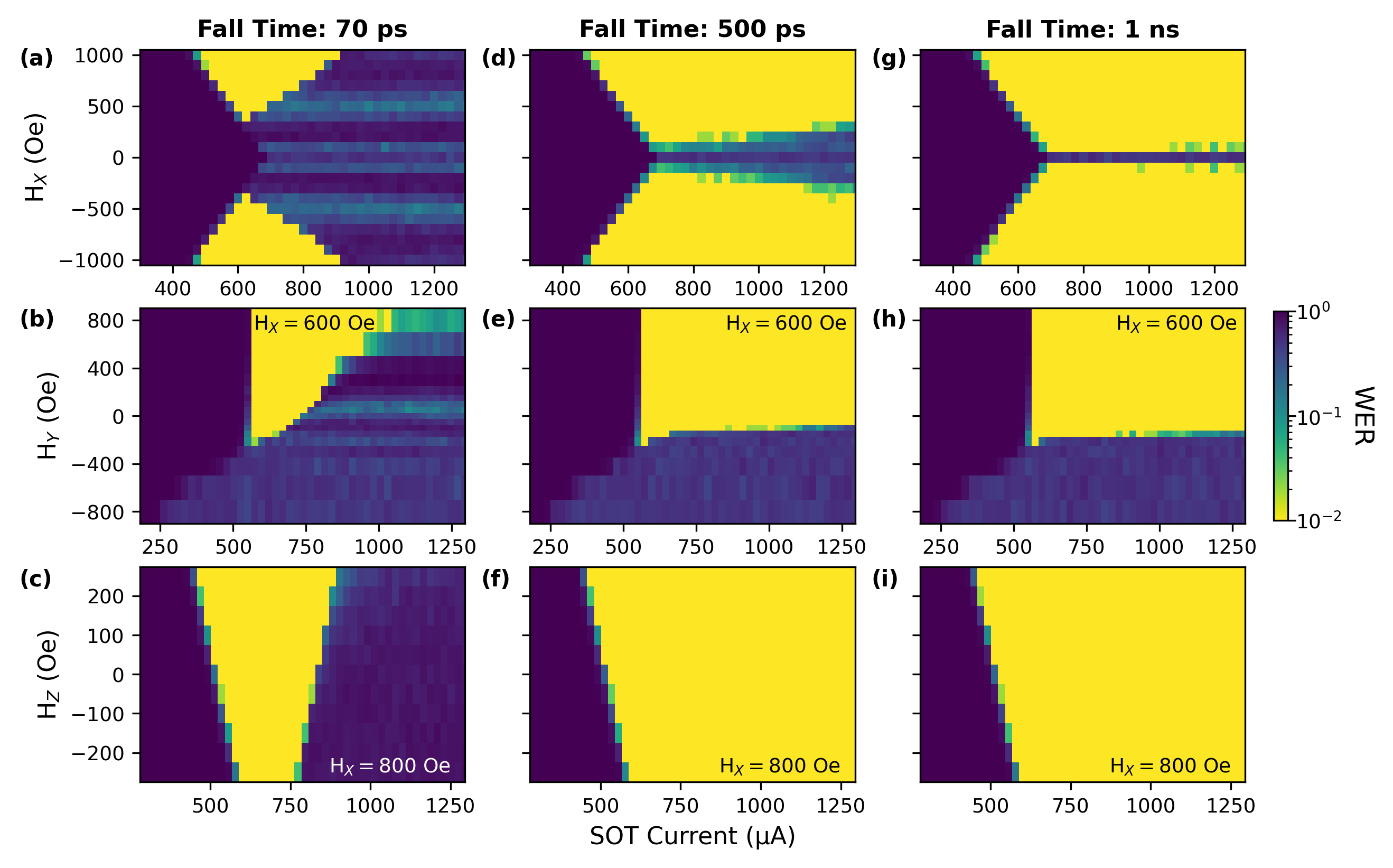}
\caption{\label{fig:s10} Simulated WER color maps as a function of applied SOT current, obtained over 100 write/read cycles, for three pulse fall times and three field configurations each: Fall time$~=70~$ps (a) variable H$_{\text{X}}$, (b) variable H$_{\text{Y}}$ under H$_{\text{X}}=600~$Oe, (c) variable H$_{\text{Z}}$ under H$_{\text{X}}=800~$Oe; Fall time$~=500~$ps (d) variable H$_{\text{X}}$, (e) variable H$_{\text{Y}}$ under H$_{\text{X}}=600~$Oe, (f) variable H$_{\text{Z}}$ under H$_{\text{X}}=800~$Oe; Fall time$~=1~$ns (g) variable H$_{\text{X}}$, (h) variable H$_{\text{Y}}$ under H$_{\text{X}}=600~$Oe, (i) variable H$_{\text{Z}}$ under H$_{\text{X}}=800~$Oe. In all panels, blue ($\text{WER}=1$) denotes switching failure in every trial, while yellow indicates deterministic switching (no writing error, $\text{WER}=10^{-2}$).}
\end{figure}

}

\end{document}